\documentclass[twocolumn]{aastex701}
\usepackage{amsmath}
\usepackage{orcidlink}
\usepackage{soul}
\newcommand{\OIIIa}{[O$\,\textsc{iii}]$~$\lambda$4959}
\newcommand{\OIIIw}{[O$\,\textsc{iii}]$~$\lambda$5007}

\newcommand{\FeII}{Fe\,{\sc ii}}
\newcommand{\Hb}{{H}$\beta$}

\newcommand{\NeIIIdblt}{[Ne$\,\textsc{iii}]$~$\lambda\lambda$3868, 3966}
\newcommand{\SII}{[S$\,\textsc{ii}]$~$\lambda$4073}
\newcommand{\CIV}{C\,{\sc iv}}

\begin{document}
\title{Spectroscopic Variability of the Broad \Hb\ Emission Line in Sloan Digital Sky Survey Quasars}
\author[0000-0001-7306-1830]{Collin M. Dabbieri}
\affiliation{Vanderbilt University, Department of Physics \& Astronomy, 6301 Stevenson Center, Nashville, TN 37235, USA}
\email{collin.m.dabbieri@vanderbilt.edu}
\author[0000-0001-8557-2822]{Jessie C. Runnoe}
\affiliation{Vanderbilt University, Department of Physics \& Astronomy, 6301 Stevenson Center, Nashville, TN 37235, USA}
\affiliation{Fisk University, Department of Life and Physical Sciences, 1000 17th Avenue N, Nashville, TN 37208, USA}
\email{jessie.c.runnoe@vanderbilt.edu}
\author[0000-0002-3719-940X]{Michael Eracleous}
\affiliation{Department of Astronomy \& Astrophysics, and Institute for Gravitation and the Cosmos, Penn State University, 525 Davey Lab, 251 Pollock Road, University Park, PA 16802, USA}
\email{mxe17@psu.edu}
\author[0000-0002-0863-1232]{Mary E. Kaldor}
\affiliation{Vanderbilt University, Department of Physics \& Astronomy, 6301 Stevenson Center, Nashville, TN 37235, USA}
\email{mary.e.kaldor@vanderbilt.edu}
\author[0000-0001-9741-2703]{Mary Ogborn}
\affiliation{Department of Astronomy \& Astrophysics, and Institute for Gravitation and the Cosmos, Penn State University, 525 Davey Lab, 251 Pollock Road, University Park, PA 16802, USA}
\email{mk05309@psu.edu}
\author[0000-0001-9806-4034]{Niana N. Mohammed}
\affiliation{Department of Astronomy \& Astrophysics, and Institute for Gravitation and the Cosmos, Penn State University, 525 Davey Lab, 251 Pollock Road, University Park, PA 16802, USA}
\affiliation{Fisk University, Department of Life and Physical Sciences, 1000 17th Avenue N, Nashville, TN 37208, USA}
\email{nnm5189@psu.edu}

\begin{abstract}

We present a catalog of broad \Hb\ variability properties for all spectra of quasars with $z<0.8$ and at least two observations included in the Sloan Digital Sky Survey (SDSS) Data Release 16 quasar catalog. For each spectrum, we perform a spectral decomposition to isolate the broad \Hb\ emission. We measure the luminosity, FWHM, equivalent width, centroid, and Pearson skewness coefficient of broad \Hb\ and provide derived physical properties such as the single-epoch black hole mass and the bolometric luminosity. For each pair of spectra in the sample, we calculate the change in radial velocity of the centroid of broad \Hb\ emission ($\Delta v_{rad}$) as well as other derived properties related to broad \Hb\ shape variability. We use forward-modeling methods to estimate the uncertainty in our measurements and discuss an improved method for estimating the uncertainty in $\Delta v_{rad}$ in the case where a spectral decomposition is used to isolate the broad \Hb\ emission. We find that $\Delta v_{rad}$ is not normally distributed and that the shape of the distribution depends on the interval between observations. We discuss the effect of the predominance of the Reverberation Mapping subsample in the sample of pairs of spectra in SDSS.

\end{abstract}


\section{Introduction}\label{intro}

Quasars are the most luminous non-exploding objects in the universe. The important regions of the central engine where the energy is generated and processed can be difficult to study because of their small sizes and typical distances. Quasar central engines are spatially unresolved in all but a few special cases \citep{gravity2018,eht2019,eht2022}. However, it is possible to probe the unresolved structure by looking at the information-rich spectra of quasars, and especially by studying the variability of their spectra over time. 

One of the most prominent features of the optical/UV spectra of quasars are the broad emission lines. The shape of the line profile for a broad emission line encodes information about the distribution, properties, and dynamics of the emitting gas, which is located in the so-called broad-line region (BLR). The BLR is a smooth \citep{arav1997,arav1998,dietrich1998,laor2006} distribution of gas that may be a disc-like extension of the accretion disc \citep[e.g.,][]{wills&browne1986,Dumont1990IV}. The radius of the emitting region is of order light-weeks, though this can vary substantially over time and for different quasars and different broad emission lines \citep[e.g.,][]{bentz2009}. Dynamically, the gas is virialized for at least some emission lines for some quasars \citep{peterson2000,wang2020}.  But BLR gas may also consist of a non-disc component (potentially a disc wind) that is preferentially visible in certain emission lines, like \CIV, over others \citep[e.g.,][]{Emmering1992,elitzur2014,elitzur2016,wang2020}. 

The properties of the broad emission lines emitted from the BLR, including their fluxes, centroids, and profile shapes, are observed to change over time. The variability is driven by different physical phenomena on different timescales. The broad lines vary on light-crossing timescales as the BLR gas responds to changes in flux from the accretion disc, in a process known as reverberation. Reverberation mapping (RM) aims to map the spatially unresolved line-emitting structure using time delays between the accretion disc and photo-ionized BLR gas \citep{blandfordmckee,peterson1992}. RM studies have produced accurate measurements of the masses of many supermassive black holes, under the assumption that the gas is virialized, making it possible to build scaling relations for the black hole mass to allow for single epoch mass estimates \citep{grier2017,grier2019,homayouni2019,shen2019,alvarez2020,bonta2020,homayouni2020,li2023,shen2024}.

The BLR also shows variability on dynamical timescales as the gas itself moves. For example, \cite{sergeev2007} analyzed 827 optical spectra of NGC 5548 over a 30 year period, and found that the \Hb\ profile can vary dramatically over periods of months to years, in ways not represented by reverberation  \citep[see also][]{wanders1996}. In particular, they found the magnitude of variation in the centroid of the line was significantly larger on long timescales than on reverberation timescales, and that the profile can completely change shape, indicating significant changes in the kinematics of the emitting gas on long timescales. 

Existing analyses of spectroscopic variability in quasars, like \cite{sergeev2007}, have focused on a small number of individual objects, due to the lack of large samples of time-domain quasar spectra. With the introduction of wide-field spectroscopic surveys, such as the Sloan Digital Sky Survey (SDSS), we have dramatically expanded the sample of known quasars with spectroscopic observations (Pre-SDSS: \citealt{schmidtgreen1983,hewett1995,croom2001}; SDSS: \citealt{schneider2003,schneider2005,schneider2007,schneider2010,paris2012,paris2014,paris2017,paris2018,lyke2020}). However, the first three iterations of SDSS were focused primarily on cosmology, and thus only targeted individual quasars for single spectroscopic observations. Thus, existing large sample analyses have focused on constraining single epoch quasar properties at the population level. Catalogs have provided single epoch model fits for spectroscopic properties of all the quasars from Data Release 7 \citep{shen2011} and Data Release 16 \citep[DR16; ][]{wushen2022}. The availability of these catalogs that take a uniform approach to estimating hundreds of important quasar properties for such a large sample of quasars allows for an unprecedented level of breadth and depth in studying quasar central engines \citep{kellyshen2013,shen2014nature,ghisellini2014}. 

SDSS is now in its fifth iteration, and the last two iterations have focused on taking repeat spectroscopy for previously observed quasars. The sample of time-domain spectra is now extensive, with many thousands of pairs of observations separated by baselines that correspond to BLR dynamical times. What still remains is to construct a large catalog of variability properties for the now-extensive sample of time-domain spectroscopy available in SDSS, akin to what \cite{wushen2022} did for single-epoch spectroscopy. 

In this work, we provide a catalog of spectroscopic variability properties for all public SDSS spectra through DR16, focused on the broad \Hb\ (b\Hb) emission line. In Section \ref{data}, we define the sample and available data. In Section \ref{decomp}, we perform a spectral decomposition in order to isolate the broad \Hb\ emission line. Section \ref{measurements} discusses the measurement of the change in radial velocity of the centroid of broad \Hb\ ($\Delta v_{rad}$) and the estimation of its uncertainty. In Section \ref{results}, we discuss the measurement of other properties of the broad line profile, including the luminosity, FWHM, equivalent width, and Pearson skewness coefficient of broad \Hb, as well as the change in these parameters for each pair of spectra. We explore the distribution of these measured variability properties of the line profile as a function of the rest-frame time between observations. In Section \ref{discussion}, we discuss some challenges with time-domain analysis of SDSS spectra, and in Section \ref{summary} we summarize our results. Throughout this work, we adopt a cosmology with $H_0=70$ km $s^{-1}$, $\Omega_\Lambda=0.7$, and $\Omega_m=0.3$.

\section{Sample and Data}\label{data}

We selected the sample of quasars with multiple epochs of spectroscopy through the fourth iteration of SDSS \citep{blanton2017}. In SDSS~I \citep{york2000} through SDSS~III \citep{eisenstein2011}, multiple epoch spectroscopy was largely serendipitous, with the exception of a small number of SDSS~IV pilot programs observed at the end of SDSS~III. Thus, spectra from these iterations are primarily used in this work as an initial epoch for objects that were re-observed in SDSS~IV. The two programs that are the main source of time-domain spectra of AGN in SDSS~IV are the Reverberation Mapping Project \citep{shen2023} and the Time Domain Spectroscopic Survey \citep[TDSS;][]{morganson2015}.

The SDSS Reverberation Mapping Project \citep{shen2016} is an observing campaign designed to make detailed spectroscopic reverberation measurements at high cadence in order to measure the time lag between the variability of the continuum and the variability of the broad emission line for a sizable sample of several hundred quasars. The RM program targets known quasars with  $i<21.7$~mag  in the 7~deg$^2$ RM field \citep{rm}. RM targets received roughly 70 spectroscopic observations sampling baselines from days to months, but RM quasars may also have earlier spectra, so a given RM quasar may have many pairs of spectra covering baselines of years to a decade.

With TDSS, SDSS~IV also implemented observing strategies to take repeat observations of a much larger number of quasars at lower cadence. These made it possible to study rare and previously unexpected phenomena in the time-domain such as changing-look quasars, AGN that transition between quasar-like and galaxy-like states or vice versa \citep{lamassa2015,runnoe2016,macleod2016,yang2018,macleod2019,guo2020,pottsvillforth2021,green2022,zeltyn2022,zeltyn2024}.

TDSS had two channels for target selection. The first channel targeted objects whose optical luminosity varied by a tenth of a magnitude or more on timescales of a year or less \citep{morganson2015}. This selection covers the vast majority of quasars as well as about one percent of stars. Roughly 2/3 of TDSS variables selected in this channel are spectroscopically classified as quasars. 90\% of TDSS targets fall under the first channel and receive single-epoch spectroscopy for identification and classification. As such, these targets are not typically included in our sample.

The second channel for TDSS target selection covers the Few Epoch Spectroscopy (FES) projects. The quasar FES projects target broad absorption line trough variability (TDSS\_FES\_VARBAL), Balmer line variability in high signal-to-noise quasars (TDSS\_FES\_NQHISN), double peaked broad emission line quasars (TDSS\_FES\_DE), searches for binary black hole quasars via \ion{Mg}{2} velocity shifts (TDSS\_FES\_MGII), and the most optically variable quasars (TDSS\_FES\_HYPQSO). The more recent TDSS Repeat Quasar Spectroscopy (RQS) program \citep{macleod2018} has obtained multi-epoch spectra for 16,500 known quasars, sampling a broad range of physical properties, reducing bias toward certain quasar subclasses compared to the FES program. 

We began the selection of our sample with all of the objects in the Sloan Digital Sky Survey DR16 Quasar Catalog \citep[DR16Q;][]{lyke2020}. This is the final catalog of SDSS~IV and provides both new observations as well as observations from previous data releases. It contains 750,414 quasars, and for each quasar, it aims to provide all SDSS spectra, including repeat observations of the same quasar, with a grand total of 950,318 spectra. Of the quasars in DR16Q, 119,924 have more than one spectral observation, with a total of 319,828 spectra for these quasars. Adding a redshift cut of z$<$0.8, which is necessary to view \Hb\ emission, truncates the sample to 31,320 spectra for 10,336 quasars with repeat spectroscopy. 

We define SNR(b\Hb), the signal to noise of broad \Hb, as the peak flux density of the broad \Hb\ data profile (after isolating the broad line profile which we perform in Section \ref{measurements}) divided by the original spectrum uncertainty at the same wavelength. To control for data spikes, we take the median \Hb\ flux density and uncertainty for the 10 adjacent wavelength bins centered on the \Hb\ peak flux. We perform a further data quality cut for spectra with SNR(b\Hb) less than 5 (after carrying out the decomposition).

In order to search for quasars that may have been left out of DR16Q we cross matched all DR16Q quasars against two different sources of SDSS spectra, a locally curated list of SDSS-I through SDSS-III quasars (priv. comm. A. Myers), and a target list for TDSS. We found 634 objects that were in SDSS I-III quasar catalogs that were not in DR16Q, and 74 TDSS quasars that were not in DR16Q. For each of these objects, if we were able to find multiple SDSS spectra, we included them in the sample.

\begin{figure}
\includegraphics[width=\linewidth]{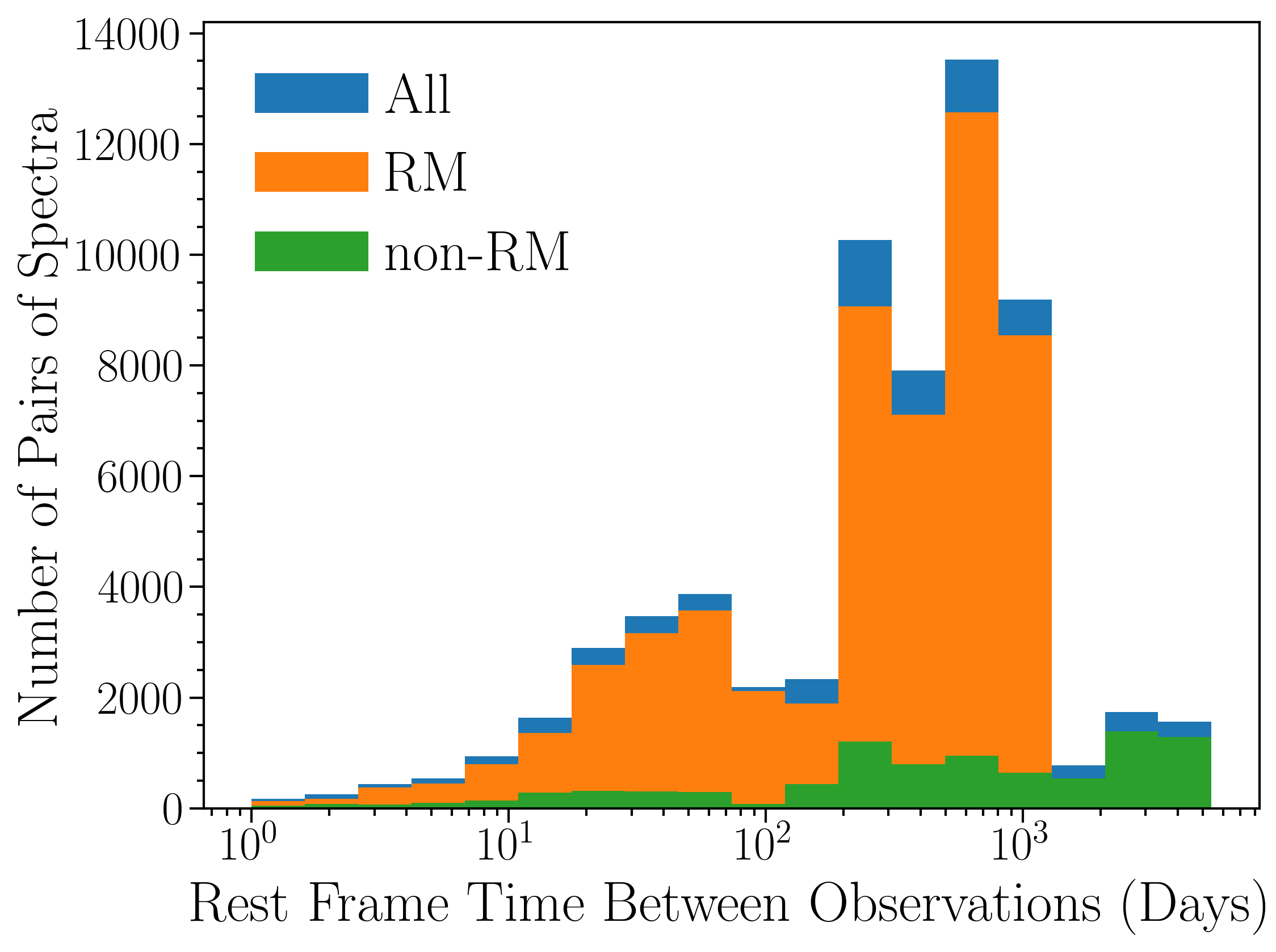}
\caption{The distribution of the pairs of spectral observations in our sample as a function of rest frame time between observations. Also plotted are the histograms for the RM and non-RM sub-samples. The RM-targeted pairs dominate the dataset for all but the longest baselines. This is the first sample large enough to statistically characterize BLR profile variability on timescales comparable to BLR dynamical times.}
\label{fig:dt_hist_rm}
\end{figure}

Fig.~\ref{fig:dt_hist_rm} shows the distribution of the sample in spectral pair space as a function of the rest-frame time between observations. Because a quasar with 70 spectra will have 2415 unique pairs of spectra, the RM sample dominates our dataset, which is measured in pairs of spectra. The non-RM sample only makes up a significant fraction of the pairs of spectra at the longest time baselines.

Spectra from SDSS I-II use the SDSS spectrograph, with a wavelength range of 3800 to 9100 \AA. Spectra from SDSS III-IV use the BOSS spectrograph \citep{smee2013}, with a wavelength range of 3600 to 10,400 \AA. Both spectrographs have a spectral resolution of $\lambda/\Delta \lambda \approx 2000$. BOSS spectra were processed with the \textsc{spec1d} pipeline \citep{bolton2012} and SDSS I-II spectra were processed with an earlier version of the same pipeline. For each spectrum, we corrected for Milky Way dust extinction \citep{ccm} and shifted to the quasar's rest frame, using DR16Q redshifts when available and pipeline redshifts when necessary. 

\section{Spectral Decomposition}\label{decomp}

In order to achieve the most accurate emission-line measurements, we first performed a spectral decomposition to isolate the broad \Hb\ emission from other components of the spectrum. We simultaneously modeled the power law continuum, Fe II complex, host galaxy continuum, Balmer continuum and emission line components of our spectra, from 3750 to 5550 \AA\ in the rest frame. 

Ideally, to obtain the best fit with a model of this many parameters, one would use forward modeling with Markov Chain Monte Carlo (MCMC) methods to avoid being trapped in local minima of the $\chi^2$ hypersurface and to provide robust uncertainty estimates for model parameters \citep{badass}. However, the size of our sample made this computationally prohibitively "expensive", so we adopted a more efficient method. This method uses a few fast $\chi^2$ optimization algorithms and then picks the result with the best $\chi^2$. 

\begin{figure*}
\includegraphics[width=\linewidth]{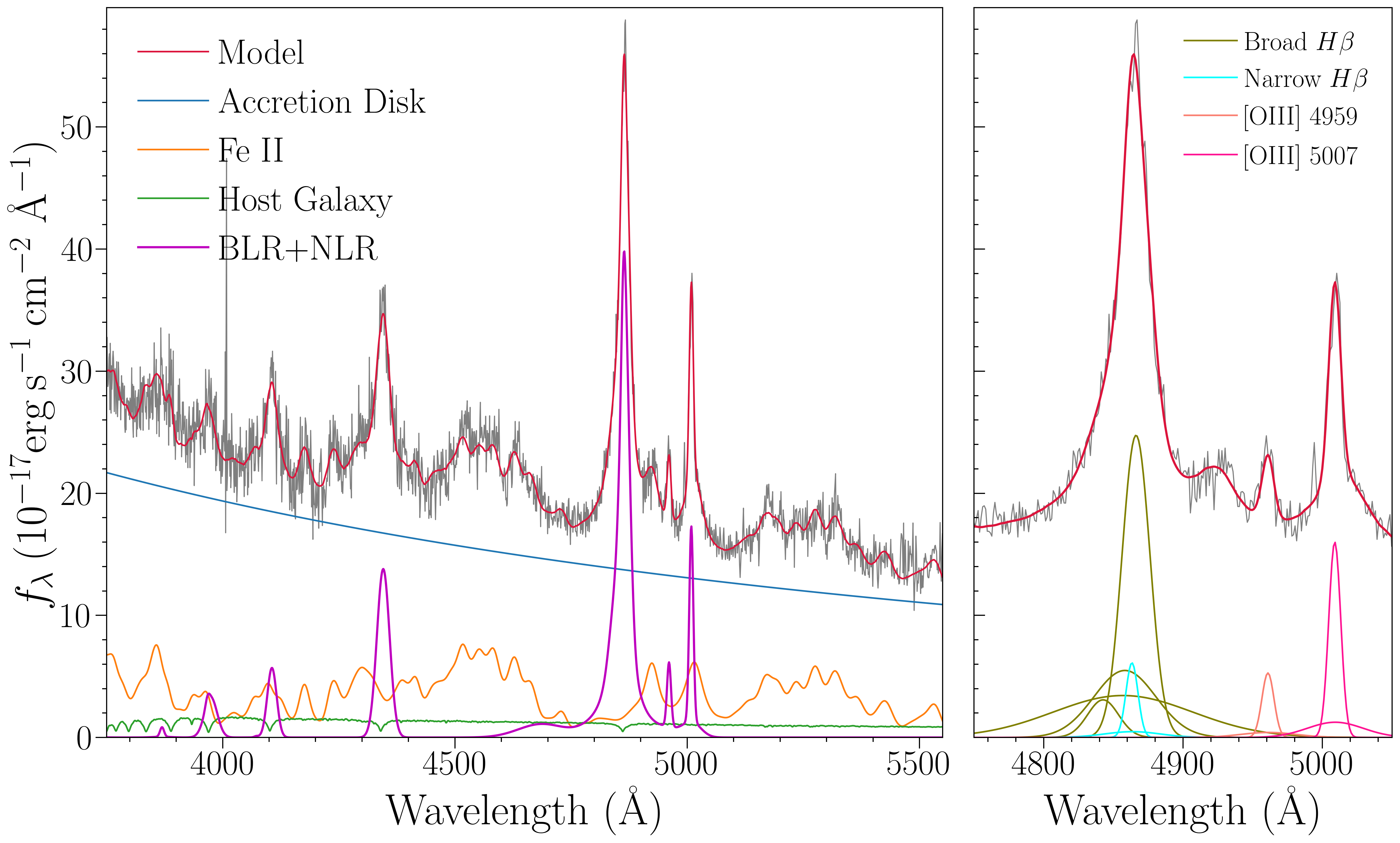}
\caption{An example spectral decomposition. Left: The data (gray) are shown with the total decomposition model (red). The model components include a power law for the accretion disk (blue), BLR and NLR emission lines (magenta), host galaxy template (green), Fe II template (orange).
Right: A zoom-in highlighting the \Hb\ region of the decomposition. The data (gray) and total model (red) are plotted along with the individual Gaussians that characterize \OIIIw\ (pink), \OIIIa\ (salmon), narrow \Hb\ (cyan), and broad \Hb\ (olive).}
\label{fig:specdecomp}
\end{figure*}

We took a number of steps in order to mitigate the local minimum problem. First, we made intelligent guesses for the initial parameters of our model (Section~\ref{init}). Second, we performed multiple fits, limiting each fit to a specific wavelength region, and holding all parameters fixed except those that are relevant to that region (Section~\ref{fitorder}). Third, we used multiple optimizers when fitting, and selected the optimizer with the best fit as the final result for each object (Section~\ref{opt}). We describe all of our model components and fitting methods below. An example fit is shown in Fig. \ref{fig:specdecomp}.

\subsection{Model Parameters}

\begin{itemize}

\item Host Galaxy - Host galaxy templates were generated from the model stellar population spectra produced by MILES \citep{miles}. The MILES templates provide model host galaxy spectra for a range of ages of the stellar population, from 30 Myr to 13 Gyr.  Our model has two free parameters for the host galaxy—the age of the stellar population, allowing for linear interpolation between ages, and a scale factor for the template. 

\item Optical \ion{Fe}{2} - Optical \ion{Fe}{2} templates were created using the I Zw I optical \ion{Fe}{2} template \citep{boroson&green} convolved with Gaussians of various widths, from 1 to 30 \AA. The model has a free parameter for the width of the convolution kernel, allowing for interpolation between convolution widths, and a scale factor for the template.


\item Balmer High Order Lines - We created a pseudocontinuum from blended, high order Balmer lines, starting with H$\epsilon$, based on the atomic parameters from \cite{storey1995}, which includes the relative strengths of the Balmer lines up to the first 50 energy levels of H \citep[see also][]{kovacevic2014}. We set the temperature to 15,000 K and the density to $10^{10}$~cm$^{-3}$ and convolved the template with Gaussians of varying Full Width at Half Max (FWHM), from 1,000 to 11,000~km s$^{-1}$, allowing for interpolation between created templates. The model is also given a scale factor for the high order templates.

\item Emission Lines - In general, we fit broad emission lines with 1-4 Gaussians, depending on the strength of the line, the complexity of its profile, and on the importance of having accurate fits for that line. We fit narrow emission lines with two Gaussians. 

Near 5000~\AA, we fit broad and narrow \Hb, and [O III] $\lambda\lambda4959,5007$. In order to constrain narrow \Hb\ emission, we restricted all of these narrow line region (NLR) emission lines to have the same line profile shapes. Each line is allowed a variable amplitude for the line profile, with the exception of \OIIIa, which is restricted to have 1/3 the flux of \OIIIw. Broad \Hb\ was fit with four Gaussians.

We fit a number of narrow lines near 4000~\AA\ to better constrain nearby host galaxy absorption lines. These lines are \NeIIIdblt, and \SII. Each line in a doublet was fit with two Gaussians, with the restriction that each line has the same line profile shape. This means that each narrow line has a single parameter for the amplitude of the line profile. The group of narrow lines collectively has one parameter for the ratio of the amplitudes of the two Gaussians, two parameters for the offset in wavelength from rest for each of the two Gaussians, and two parameters for the width of the two Gaussians. We also fit broad H$\delta$ and broad H$\gamma$ with a single Gaussian for each.

We restricted the narrow lines near 4000~\AA\ to have the same line profile shapes and we restricted the narrow lines near 5000~\AA\ to have the same line profile shapes. These two groups of narrow lines were handled separately not for physical reasons, but because our fitting procedure (Section \ref{fitorder}) has separate fitting steps for these two wavelength regions.

\end{itemize}

\subsection{Initial Parameter Estimates}\label{init}

In order to alleviate the local minimum problem, we automatically made intelligent guesses for some initial parameter values before modeling. First, we checked the flux at blue and red continuum wavelengths, 4000~\AA\ and 5500~\AA\ respectively, and picked power law parameters that pass through those two points. Next, we checked the flux at the rest wavelength of each of our emission lines, and used the difference in flux between the power law and the spectrum at that wavelength to pick initial amplitude values for our emission lines.

\subsection{Model Fitting Procedure}\label{fitorder}


In order to fit a model with so many free parameters, it was necessary to perform multiple fits on subsets of model parameters, holding other parameters fixed. This allowed the dimensionality of the parameter space of each fit to remain tractable for fast-fitting $\chi^2$ optimization methods. It also allowed us to place priorities for the quality of the fit in different parts of the spectrum. For our first fit, we restricted the data to selected continuum wavelength regions of 4150-4250 \AA, 4450-4800 \AA\ and 5050-5550 \AA\ and fit only power law and \FeII\ model parameters. Next, we restricted the data to wavelengths of 3750 to 4080 \AA\ and fit parameters for the host galaxy, the narrow emission lines near 4000~\AA, and the Balmer continuum. For the third fit, we restricted the data to 4000-4500 \AA\ and fit the H$\delta$ and H$\gamma$ emission lines. For the fourth fit we restricted the data to 4600-5100 \AA\ and fit parameters for broad \Hb, narrow \Hb, \OIIIa, and \OIIIw. At this point all model parameters had been fit once and the model should have a good solution. For the final fit, we used the parameter values from the previous fits as the starting point and fit all of the data and allowed all parameters to vary. This procedure enabled the algorithm to make slight alterations to optimize the fit.

\subsection{Optimizers}\label{opt}

We used the python package {\tt lmfit} to fit our model parameters. One of the advantages of using this package is that it allows for the simple use of many different optimizers. We tested each {\tt lmfit} optimizer on small samples of spectra and selected the best optimizers for our model based on the quality of the fit and the computational cost. We chose the BFGS, Nelder, and Powell optimizers for our final fits. For each spectrum, the spectra were fit with each optimizer, and the result with the best $\chi^2$ was selected as the final fit. The Powell, Nelder and BFGS optimizers were selected 28074, 2969, and 37 times respectively.

\section{Measurements and Uncertainties}\label{measurements}

The results of the spectral decomposition described in Section~\ref{decomp} were used to measure properties of the broad \Hb\ line profile, and the differences of these properties between epochs. Our measured properties include the change in the centroid of broad \Hb, the change in the 5100 \AA\ continuum luminosity, the change in the integrated luminosity of broad \Hb, and the change in the FWHM of broad \Hb. Some of these properties were measured using the broad \Hb\ data profile, which was created by subtracting out every model component from the spectrum except the broad \Hb\ model component. Other properties were measured via the broad \Hb\ model profile itself, in order to reduce the sensitivity to noise. 

\subsection{Measurement of Radial Velocity Change by Cross-Correlation}\label{jitter}

We measured the change in centroid of broad \Hb\ ($\Delta v_{rad}$, also known as jitter) by calculating the cross-correlation function for the two broad \Hb\ data profiles in the pair and finding the shift (in \AA\ or equivalently km s$^{-1}$) at which the two profiles best overlap. Historically, this has been done for small samples of pairs of spectra using the full spectrum without first performing a spectral decomposition \citep{eracleous2012}. Because other components of the spectra (predominantly narrow \Hb\ emission and \FeII) have flux at the same wavelength as broad \Hb, this method requires visual inspection of the cross-correlation to ensure it finds the correct result. Following \cite{shen2013}, we instead performed a spectral decomposition first and then calculated the cross-correlation on the broad \Hb\ data profiles. 

For each step in the cross-correlation, corresponding to a specific velocity shift, we shifted one broad \Hb\ profile according to that velocity, re-binned both profiles onto a common wavelength vector using a method that preserves the total flux, and normalized the profiles by applying a scale factor to one of them in order to match the flux density of the line peak. We then calculated the $\chi^2$ for the two normalized spectra. The change in centroid of the broad emission line is given by the lowest $\chi^2$ value. We calculated this by fitting a parabola to the 5 lowest points on the $\chi^2$ curve.

\subsection{Uncertainty in $\Delta v_{rad}$ } \label{error}  

The standard method for estimating the uncertainty in $\Delta v_{rad}$, $\sigma(\Delta v_{rad})$, is to take the widest $\Delta v_{rad}$ for which $\chi^2(\Delta v_{rad}) = \chi^2_{min}+\Delta \chi^{2}$ on the cross-correlation curve \citep{eracleous2012}. A $\Delta \chi^{2}$ value of 1 represents a 1$\sigma$ deviation for a $\chi^2$ distribution with 1 degree of freedom \citep{lampton1976}. However, we first performed a spectral decomposition in order to isolate broad \Hb\ emission, and then calculated the cross correlation curve for the broad \Hb\ data profile. Therefore our uncertainties in $\Delta v_{rad}$ must take into account both the uncertainty from the spectral decomposition and the uncertainty in the cross-correlation.

In order to properly calculate the uncertainties for the spectral decomposition, we fit a pilot sample with MCMC, as it provides more robust fits and uncertainty estimates for model parameters. We selected the pilot sample by making a grid in rest-frame time between observations and SNR(b\Hb) and randomly selecting a pair of spectra from each gridpoint, if one such pair exists. This gave us a sample of 170 pairs that evenly sample the parameter space. We used this pilot sample to construct scaling relationships for the uncertainties based on easily measured properties, such as SNR(b\Hb). 

We performed the MCMC spectral decompositions for our pilot sample with the affine-invariant sampler {\tt emcee} \citep{emcee}. We fit each pair in the sample using the model described in Section \ref{decomp} but replacing the fitting order and optimizer scheme described in Sections \ref{fitorder} and \ref{opt} with an {\tt emcee} fit. We took the chain and discarded the burn-in, thinned by the autocorrelation length, and flattened the {\tt emcee} walkers to provide independent samples of the posterior distribution for our model parameters.

We performed a Monte Carlo simulation for $\sigma(\Delta v_{rad})$ by taking independent samples from the posterior distribution of the {\tt emcee} fits for each pair of spectra. Each sample from the posterior distribution of the model fit gave us one realization of a broad \Hb\ data profile that we used to calculate a point estimate and uncertainty in $\Delta v_{rad}$ using cross-correlation and the $\chi^2_{min}+1$ method for the uncertainty. We combined the $\Delta v_{rad}$ distributions from each sample of the {\tt emcee} posterior into a single $\Delta v_{rad}$ distribution for each pair of spectra in the pilot sample.

We then looked for quickly-measured parameters of the spectra that could serve as independent variable(s) to predict the uncertainty produced by a full MCMC fit. We tested four different terms for their significance in predicting $\sigma(\Delta v_{rad})$: $\Delta v_{rad}$, SNR(b\Hb), Fe II strength, and FWHM(b\Hb). $\Delta v_{rad}$ was chosen because one might expect larger absolute uncertainties associated with lines that shift more. SNR(b\Hb) was chosen because it is sensitive to the prominence of the broad \Hb\ line. The FWHM is sensitive to how sharp the emission line is and therefore may impact the comparison with another line profile in the cross-correlation, and Fe II strength can impact the broad \Hb\ line via imperfect modeling and subtraction. The Fe II strength did not have a significant correlation with $\sigma(\Delta v_{rad})$ at 95\% confidence, so it was disregarded. We then fit a multiple linear regression (MLR) to $\sigma(\Delta v_{rad})$ with independent variables of SNR(b\Hb) and FWHM(b\Hb), but the FWHM term was not significant at the 95\% confidence level according to the student-t test, so it was disregarded. Finally, we similarly fit an MLR with independent variables SNR(b\Hb) and $\Delta v_{rad}$. In this fit, all terms were significant at 95\% confidence. Additionally, this estimate agreed most closely with the independent estimate for the uncertainty discussed in Section \ref{lowdt}, so we adopt this estimate throughout. A binned representation of the MLR fit is provided in Fig. \ref{fig:lowdt}. The analytic form for the uncertainty is $\sigma(\Delta v_{rad})=-3.5$SNR(b\Hb)+$0.31|\Delta v_{rad}|+91$ for values of SNR(b\Hb) between 0 and 40 and values of $\Delta v_{rad}$ between -1500 and 1500 km s$^{-1}$. 


These efforts to characterize the uncertainties on the properties of the broad \Hb\ line also revealed that this approach is very sensitive to some details of the method. The different fits discussed above produce significantly different uncertainty estimates. In this sense, the uncertainty estimate is very sensitive to the model selection. Having an independent estimate of the uncertainty to compare to was valuable.

\subsection{Estimating the Uncertainty in \Hb\ $\Delta v_{rad}$ with short baseline pairs} \label{lowdt}

\begin{figure}
\centering
\includegraphics[width=\linewidth]{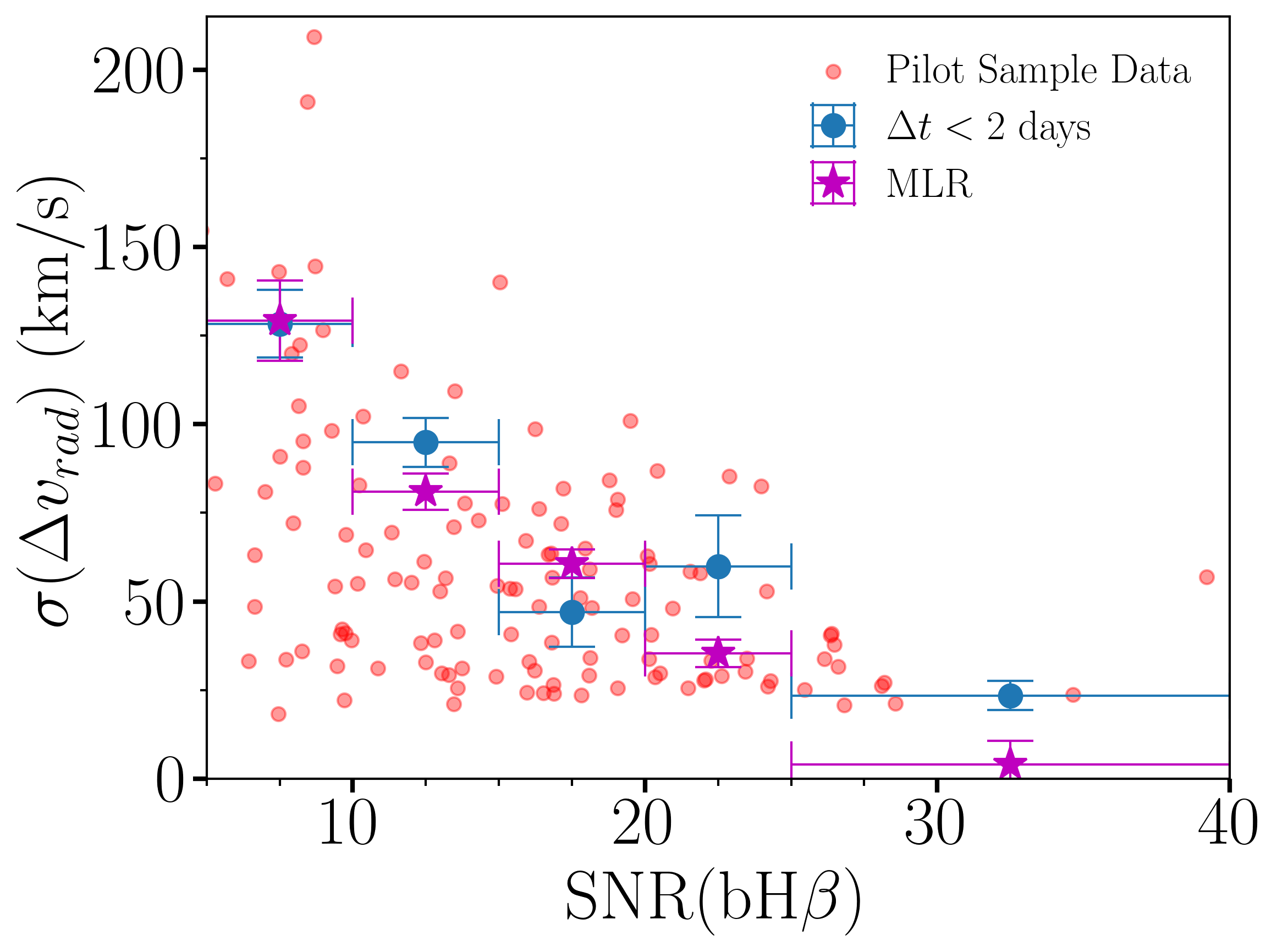}
\caption{Estimates for the uncertainty in radial velocity jitter, $\sigma(\Delta v_{rad})$. The multiple linear regression fit of $\sigma(\Delta v_{rad})$ with SNR(b\Hb) and $\Delta v_{rad}$ as the independent variables is plotted in purple. In order to represent the fit on a single independent variable, the sample was binned by SNR(b\Hb) and the mean uncertainty value for each bin is plotted. Uncertainty estimates from the $\Delta t<2$ days sample are plotted in blue. For both the MLR and $\Delta t<2$ days plots, horizontal error bars represent the binning and the vertical error bars are estimated with the bootstrap method. The $\Delta t<2$ days bins include 166, 93, 21, 34, and 12 pairs per bin respectively from low to high SNR. }
\label{fig:lowdt}
\end{figure}

As a test of our method for estimating the uncertainty in \Hb\ $\Delta v_{rad}$, we also calculated a second estimate for the same uncertainty. As argued in \cite{kozlowski2017}, as the time between a pair of observations approaches zero, the true variability approaches zero. Therefore one can take pairs of observations with very short baselines and use the observed variability as a noise estimate. Following \cite{kozlowski2017}, we took all pairs with $\Delta t<2$ days as our short baseline sample, returning 326 pairs of spectra. For each pair, we performed a spectral decomposition as described in Section~\ref{decomp} and calculated the \Hb\ jitter using the standard method described in Section~\ref{jitter}. Each jitter measurement for the low $\Delta t$ sample represents a single draw from the noise distribution for $\Delta v_{rad}$. To compare the low $\Delta t$ estimate of $\sigma(\Delta v_{rad})$ to our pilot sample uncertainty estimate (Section \ref{error}), we binned the sample as a function of SNR(b\Hb) and calculated the standard deviation of $\Delta v_{rad}$ for the low $\Delta t$ pairs in each bin. We estimated the uncertainty on $\sigma(\Delta v_{rad})$ in each bin using the bootstrap method. Fig. \ref{fig:lowdt} shows the pilot sample uncertainties in red with the MLR fit for the uncertainty in purple and the $\Delta t<2$ days sample estimates for the uncertainty in blue.


In measurements of ensemble uncertainty, we argue that it is preferable to use the low $\Delta t$ uncertainty estimate, as it is only sensitive to the hyperparameters that go into the model fitting for the spectral decomposition, and does not require fitting further scaling relations to pilot sample uncertainties. But the low $\Delta t$ uncertainty estimate does not provide uncertainties for individual pairs of spectra, so we report uncertainties based on the MLR regression to the MCMC pilot sample for all of the individual spectra and pairs of spectra in our sample.

\subsection{Estimating the Uncertainty for Other Observables}

We used a similar method to that described in Section~\ref{error} to estimate the uncertainty for our other measured properties, given in Table \ref{spectable}. We used all of the pairs of spectra in our MCMC pilot sample. For each pair of spectra in the sample, we took independent draws from the posterior distribution of the model parameters for both spectra in the pair. For each draw, we calculated each measured property. We used the distribution of calculated properties across samples of the MCMC posterior distribution to estimate the uncertainty for each observable. We then fit a multiple linear regression to the uncertainty for each value as a function of SNR(b\Hb) and the value itself for our MCMC pilot sample, and used the fit as the uncertainty estimate for the whole sample. We propagated the uncertainty where necessary, for example through the subtraction for the change in measured properties for each pair and for properties that are functions of our observables, like the black hole mass.

\section{Results}\label{results}

The measured properties for the spectra in our sample are given in Table \ref{spectable}. The columns are as follows:

\begin{itemize}
\item Column 1: SDSS J name
\item Column 2: SDSS spectroscopic redshift \citep[from DR16Q;][]{lyke2020}.
\item Column 3: SNR(b\Hb), calculated by taking the median SNR of the 10 points adjacent to the peak of the broad \Hb\ data profile, divided by the median spectrum uncertainty over the same wavelength bins.
\item Column 4: 5100 \AA\ continuum luminosity, calculated from the flux of the model power law at 5100 \AA.
\item Column 5: Luminosity of broad \Hb, calculated by integrating the broad \Hb\ data profile, from 4801 to 4921 \AA.
\item Column 6: FWHM of broad \Hb, calculated from the broad \Hb\ model profile. 
\item Column 7: Equivalent Width of broad \Hb. Calculated according to 

$$ EW=\sum \frac{f_{i,H\beta}}{f_{i,pl}}\delta \lambda_i $$

where $f_{i,H\beta}$ is the flux density of the isolated b\Hb\ line, $f_{i,pl}$ is the flux density of the power-law continuum, and $\delta \lambda_i$ is the width of the pixel.

\item Column 8: Velocity offset of the centroid of the broad \Hb\ model profile. The location of the centroid was calculated according to
\begin{equation}\label{centroid}
 \langle \lambda \rangle = \frac{\sum f_{\lambda,i} \lambda_i}{\sum f_{\lambda,i}} .
\end{equation}

\item Column 9: Pearson skewness coefficient, $Sk_2=3(\langle \lambda \rangle-median)/\sigma$, where $\langle \lambda \rangle$ is the centroid from Eq. \ref{centroid} and median is the median wavelength, i.e. the wavelength with equal integrated flux on its blue and red sides. We found that the Pearson skewness coefficient was sensitive to noise at the wings of the line profile, so we calculated this metric over the wavelength range that corresponds to flux above 7\% of the max flux of the broad \Hb\ data profile \citep{runnoe2015} .

\item Column 10: Black hole mass, calculated using the relation from \cite{bonta2020}:
\begin{equation}
\begin{split}
\textrm{log }(M_{BH}/M_{\odot})= \textrm{log } f + 7.015 \\
                      + 0.784(\textrm{log}(L_{bH\beta}/\textrm{erg s}^{-1})-42)\\
                      +1.387(\textrm{log }(FWHM_{bH\beta}/\textrm{km s}^{-1}) - 3.5),
\end{split}
\end{equation}
where $L_{H\beta}$ is the integrated $H\beta$ luminosity, FWHM$_{H\beta}$ is the \Hb\ full width at half maximum, and $f$ is given by $\langle \log f \rangle=0.683 \pm 0.150$ \citep{batiste2017}.
\item Column 11: Bolometric luminosity, calculated using the relation from \cite{runnoe2012}:

\begin{equation}
\begin{split}
\textrm{log }(L_{iso}/\textrm{erg s}^{-1})=(4.89\pm 1.66) \\
                      +(0.91\pm0.04)\textrm{log} (\lambda L_\lambda^{5100\textrm{\AA}}/\textrm{erg s}^{-1}).
\end{split}
\end{equation}
where $\lambda L_\lambda^{5100\textrm{\AA}}$ is the monochromatic luminosity at 5100 \AA
\item Column 12: Eddington Fraction.
\end{itemize}

\begin{deluxetable*}{lccr@{$\pm$}lr@{$\pm$}lr@{$\pm$}lr@{$\pm$}lr@{$\pm$}lr@{$\pm$}lr@{$\pm$}lr@{$\pm$}lr@{$\pm$}l}
\tablecolumns{28}
\tablewidth{\linewidth}
\tablecaption{Spectral Properties}
\tablehead{
\colhead{SDSS Name} & 
\colhead{z} & 
\colhead{SNR(H$\beta$)} & 
\multicolumn{2}{c}{log$L_{5100}$} & 
\multicolumn{2}{c}{log$L_{\rm H\beta}$} & 
\multicolumn{2}{c}{FWHM} & 
\multicolumn{2}{c}{EW} & 
\multicolumn{2}{c}{$v_{\rm cent}$} & 
\multicolumn{2}{c}{Sk$_2$} & 
\multicolumn{2}{c}{log $M_{\rm BH}$} & 
\multicolumn{2}{c}{log $L_{\rm bol}$} & 
\multicolumn{2}{c}{$L/L_{\rm Edd}$} \\
\colhead{(1)} & 
\colhead{(2)} & 
\colhead{(3)} & 
\multicolumn{2}{c}{(4)} & 
\multicolumn{2}{c}{(5)} & 
\multicolumn{2}{c}{(6)} & 
\multicolumn{2}{c}{(7)} & 
\multicolumn{2}{c}{(8)} & 
\multicolumn{2}{c}{(9)} & 
\multicolumn{2}{c}{(10)} & 
\multicolumn{2}{c}{(11)} & 
\multicolumn{2}{c}{(12)} \\
\colhead{} & 
\colhead{} & 
\colhead{} & 
\multicolumn{2}{c}{[erg s$^{-1}$ \AA$^{-1}$]} & 
\multicolumn{2}{c}{[erg s$^{-1}$]} & 
\multicolumn{2}{c}{[km s$^{-1}$]} & 
\multicolumn{2}{c}{[\AA]} & 
\multicolumn{2}{c}{[km s$^{-1}$]} & 
\multicolumn{2}{c}{} & 
\multicolumn{2}{c}{[$M_{\odot}$]} & 
\multicolumn{2}{c}{[erg s$^{-1}$]} & 
\multicolumn{2}{c}{}
}
\startdata
132850.45+651808.7&0.31&12.1&40.57&0.03&42.39&0.02&2690&130&62&38&140&20&$-$0.12&0.04&7.9&0.2&45.18&0.05&0.15&0.06\\
134719.40+590232.7&0.77&18.6&42.41&-0.03&44.32&0.01&3590&130&68&33&60&10&$-$0.17&0.04&9.6&0.2&46.85&0.04&0.15&0.06\\
023112.11$-$002056.2&0.64&11.5&40.89&0.02&42.87&0.02&4480&170&82&26&$-$130&10&$-$0.04&0.03&8.6&0.2&45.47&0.04&0.06&0.02\\
082758.52+233302.9&0.3&28.6&41.36&-0.02&43.41&0.01&4070&110&93&15&0&0&$-$0.08&0.02&9.0&0.2&45.9&0.03&0.07&0.02\\
010002.32+001642.4&0.78&6.5&41.81&0.01&43.61&0.02&4970&190&52&46&$-$170&20&0.05&0.02&9.2&0.2&46.31&0.02&0.1&0.03\\
143010.90+515406.2&0.46&7.8&40.25&0.05&42.19&0.02&2560&140&80&29&200&20&$-$0.11&0.05&7.7&0.2&44.89&0.07&0.12&0.05\\
125543.36+350952.7&0.47&11.7&41.21&0.02&43.07&0.02&7810&220&59&40&80&20&0.01&0.02&9.1&0.2&45.76&0.02&0.04&0.01\\
110057.71$-$005304.5&0.38&13.0&40.83&0.02&42.64&0.02&6570&200&52&44&190&20&$-$0.07&0.03&8.6&0.2&45.42&0.03&0.05&0.02\\
231235.79$-$083138.3&0.41&6.9&40.69&0.04&42.45&0.02&5310&190&49&47&$-$50&20&$-$0.01&0.03&8.4&0.2&45.29&0.06&0.07&0.03\\
114450.19+502609.3&0.25&16.3&40.6&0.02&42.61&0.01&5320&170&86&23&$-$50&10&$-$0.02&0.02&8.5&0.2&45.21&0.03&0.04&0.02\\
212617.68+002228.9&0.44&21.0&40.94&0.01&42.74&0.01&1800&100&58&38&10&10&$-$0.15&0.04&7.9&0.2&45.52&0.01&0.31&0.1\\
102119.95+185213.8&0.36&19.7&41.0&0.01&42.87&0.01&2970&120&66&34&30&10&$-$0.15&0.04&8.3&0.2&45.57&0.01&0.14&0.04\\
013253.61$-$015126.3&0.77&10.4&41.15&0.02&43.05&0.02&4210&160&67&35&$-$70&10&0.01&0.02&8.7&0.2&45.71&0.03&0.08&0.03\\
100613.75+210437.1&0.26&11.4&40.27&0.04&42.16&0.02&4210&160&70&34&$-$180&10&$-$0.07&0.03&8.0&0.2&44.91&0.06&0.07&0.03\\
101212.79+164009.4&0.27&13.9&40.24&0.04&42.26&0.02&3180&140&94&19&140&20&$-$0.08&0.03&7.9&0.2&44.88&0.05&0.08&0.03\\
\enddata
\tablecomments{This table is available in its entirety in machine-readable form. Square brackets indicate the units for the quantity inside the log. e.g. log$L_{5100}$[erg s$^{-1}$ \AA$^{-1}$] is the same as log($L_{5100}$/erg s$^{-1}$ \AA$^{-1}$).} 
\end{deluxetable*}
\label{spectable}

The changes in measured properties for the pairs of spectra in our sample are given in Table \ref{pairtable}. All values are calculated by subtracting the value of the chronologically earlier observation from the value of the later observation.

\begin{deluxetable*}{lccr@{$\pm$}lr@{$\pm$}lr@{$\pm$}lr@{$\pm$}lr@{$\pm$}lr@{$\pm$}lr@{$\pm$}lr@{$\pm$}l}
\tablecolumns{21}
\tablewidth{\linewidth}
\tablecaption{Pair Properties}
\tablehead{
\colhead{SDSS Name} & 
\colhead{z} & 
\colhead{$\Delta$MJD} & 
\multicolumn{2}{c}{$\Delta v_{\rm rad}$} & 
\multicolumn{2}{c}{$\Delta$ log $L_{5100}$} & 
\multicolumn{2}{c}{$\Delta$ log $L_{\rm H\beta}$} & 
\multicolumn{2}{c}{$\Delta$FWHM} & 
\multicolumn{2}{c}{$\Delta$EW} & 
\multicolumn{2}{c}{$\Delta v_{\rm peak}$} & 
\multicolumn{2}{c}{$\Delta v_{\rm cent}$} & 
\multicolumn{2}{c}{$\Delta$Sk$_2$} \\
\colhead{(1)} & 
\colhead{(2)} & 
\colhead{(3)} & 
\multicolumn{2}{c}{(4)} & 
\multicolumn{2}{c}{(5)} & 
\multicolumn{2}{c}{(6)} & 
\multicolumn{2}{c}{(7)} & 
\multicolumn{2}{c}{(8)} & 
\multicolumn{2}{c}{(9)} & 
\multicolumn{2}{c}{(10)} & 
\multicolumn{2}{c}{(11)} \\
\colhead{} & 
\colhead{} & 
\colhead{[days]} & 
\multicolumn{2}{c}{[km s$^{-1}$]} & 
\multicolumn{2}{c}{[erg s$^{-1}$ \AA$^{-1}$]} & 
\multicolumn{2}{c}{[erg s$^{-1}$]} & 
\multicolumn{2}{c}{[km s$^{-1}$]} & 
\multicolumn{2}{c}{[\AA]} & 
\multicolumn{2}{c}{[km s$^{-1}$]} & 
\multicolumn{2}{c}{[km s$^{-1}$]} & 
\multicolumn{2}{c}{}
}
\startdata
141419.84+533815.3&0.17&342&80&60&0.05&0.05&0.04&0.02&280&170&$-$1&76&$-$70&160&$-$20&10&0.06&0.06\\
140759.06+534759.7&0.17&1342&50&80&$-$0.0&0.08&$-$0.05&0.03&$-$1250&220&$-$6&66&$-$70&220&$-$110&30&0.06&0.04\\
141202.87+522026.0&0.42&508&90&80&0.04&0.05&0.02&0.02&$-$140&170&$-$1&58&0&170&$-$50&20&$-$0.01&0.07\\
140759.06+534759.7&0.17&734&70&90&$-$0.02&0.08&$-$0.08&0.03&$-$210&250&$-$4&74&$-$140&250&$-$110&30&0.04&0.04\\
141107.29+011345.3&0.63&233&$-$50&70&$-$0.11&0.03&$-$0.07&0.03&$-$70&170&4&58&0&190&0&30&$-$0.08&0.07\\
141202.87+522026.0&0.42&823&70&80&0.02&0.05&0.03&0.02&$-$70&180&3&64&0&190&$-$40&20&$-$0.01&0.08\\
141202.87+522026.0&0.42&16&60&80&$-$0.03&0.06&0.05&0.03&140&180&9&68&$-$70&190&0&30&0.04&0.08\\
121529.98+441120.3&0.54&1&80&90&$-$0.06&0.06&$-$0.06&0.03&140&210&$-$1&49&0&210&50&40&0.01&0.04\\
141625.70+535438.5&0.26&1225&40&20&0.12&0.03&0.11&0.01&340&170&$-$2&6&0&110&$-$20&10&0.01&0.0\\
141004.27+523140.9&0.53&36&30&60&$-$0.02&0.05&$-$0.1&0.03&$-$210&200&$-$11&54&70&210&30&20&0.21&0.02\\
140759.06+534759.7&0.17&310&60&80&$-$0.01&0.08&0.05&0.03&$-$420&220&6&69&$-$280&230&$-$10&30&0.04&0.03\\
141419.84+533815.3&0.17&335&30&40&$-$0.01&0.04&$-$0.1&0.02&140&160&$-$8&73&$-$70&130&$-$60&10&$-$0.03&0.06\\
142135.90+523138.9&0.25&599&660&270&0.15&0.08&0.36&0.03&890&200&17&77&$-$690&210&$-$250&40&$-$0.16&0.04\\
140508.60+530539.0&0.58&664&$-$80&60&0.05&0.02&0.09&0.02&$-$480&210&6&45&70&160&40&20&$-$0.05&0.02\\
142103.52+515819.4&0.26&21&$-$80&90&$-$0.01&0.08&0.06&0.03&$-$140&220&8&63&0&210&10&30&0.04&0.05\\
\enddata
\tablecomments{This table is available in its entirety in machine-readable form. Square brackets indicate the units for the quantity inside the log. e.g. $\Delta$log$L_{5100}$[erg s$^{-1}$ \AA$^{-1}$] is the same as $\Delta$log($L_{5100}$/erg s$^{-1}$ \AA$^{-1}$).} 
\end{deluxetable*}
\label{pairtable}

\begin{figure}
\centering
\includegraphics[width=\linewidth]{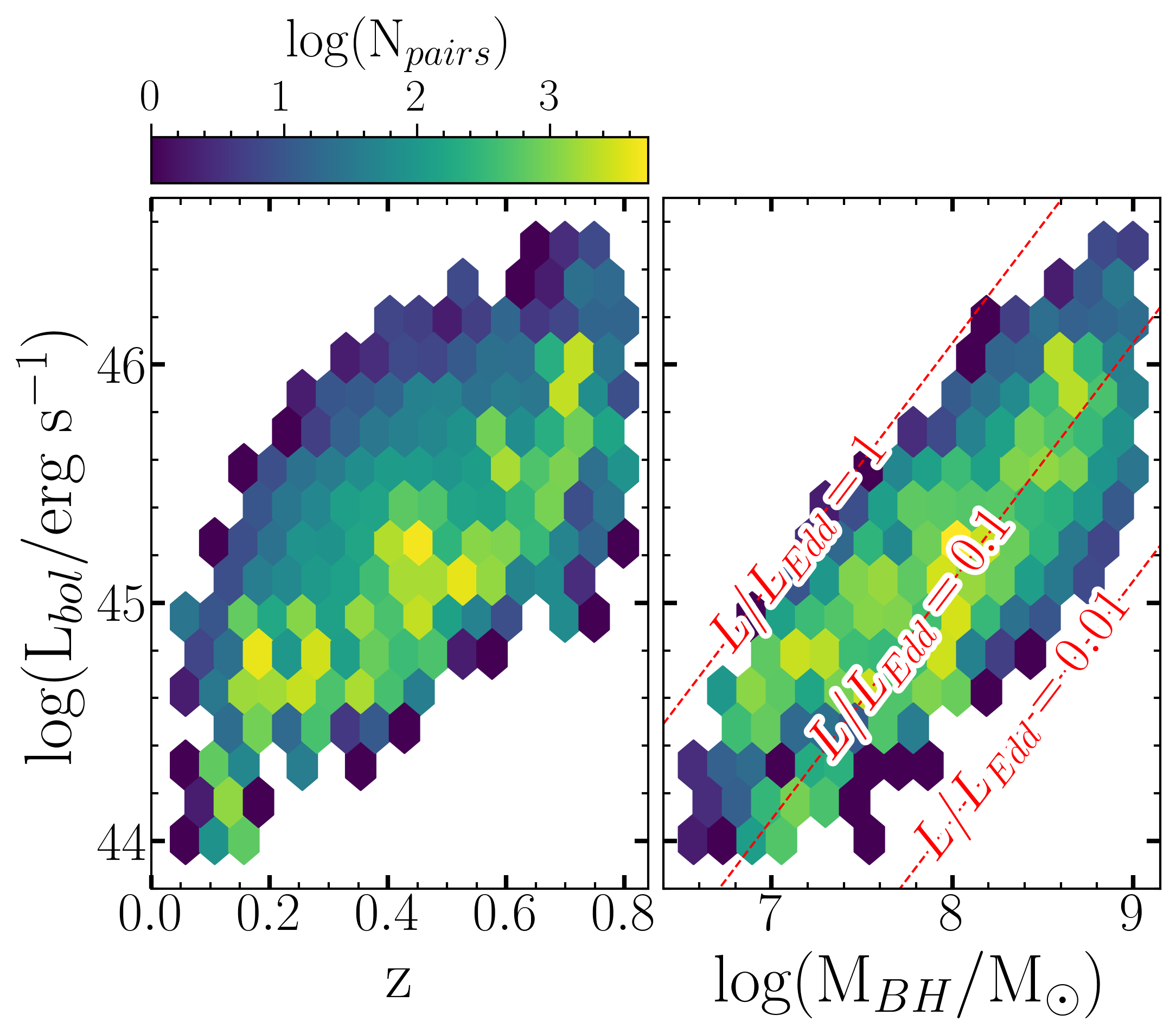}
\caption{Left panel shows the number of pairs of spectra in our dataset on a grid of bolometric luminosity and redshift. Right panel shows the same on a grid of bolometric luminosity and black hole mass. Also plotted are lines of constant Eddington ratio.}
\label{fig:lbol_mass_npairs}
\end{figure}

Fig.~\ref{fig:lbol_mass_npairs} shows two-dimensional histograms for the pairs of spectra in bolometric luminosity, black hole mass and redshift space. While the physical parameter space of the DR16Q sample is well known, variability analysis is done in pair space, and the physical properties of the sample in pair space are significantly different. For each pair of spectra, the chronologically earlier spectrum was used to estimate each physical property. We find black hole masses roughly between $10^7$ and $10^{9}$ M$_\odot$ with bolometric luminosities roughly between $10^{44}$ and $10^{47}$ erg s$^{-1}$, which is broadly consistent with other measurements of SDSS quasar spectra for a low redshift sample \citep{wushen2022}.

\begin{figure}
\centering
\includegraphics[width=\linewidth]{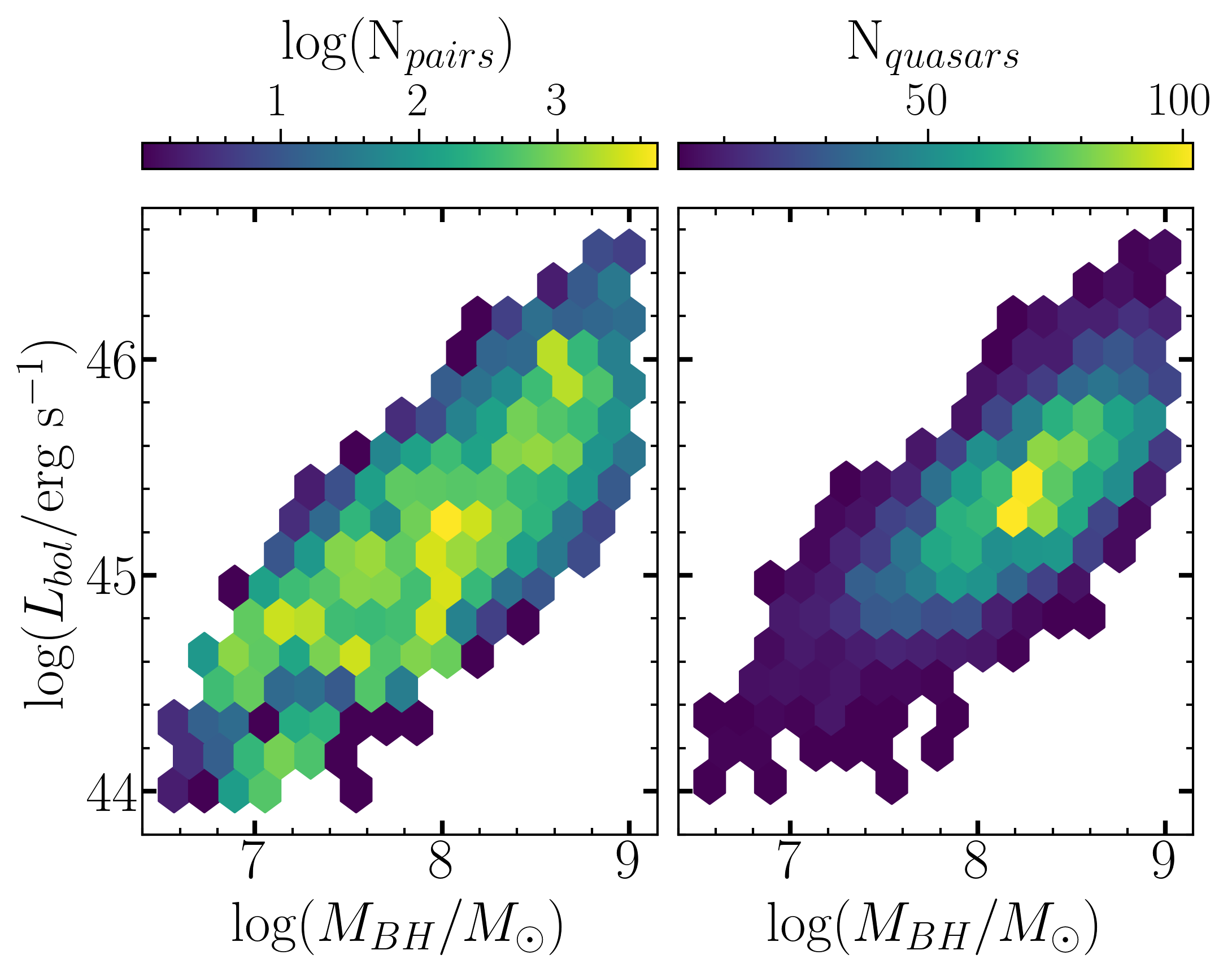}
\caption{Left panel shows the number of pairs of spectra in our sample on a grid of bolometric luminosity and black hole mass. Right panel shows the number of quasars in our sample on the same grid. Because the RM subsample makes up the vast majority of pairs of spectra in our sample and a very small fraction of quasars in our sample, this plot also illustrates the differences in physical properties of the RM and non-RM sub-samples.}
\label{fig:npairs_vs_nquasars}
\end{figure}

We plot a two-dimensional histogram for the number of pairs of spectra as a function of bolometric luminosity and black hole mass, as well as a two-dimensional histogram of the number of quasars in the same space in Fig.~\ref{fig:npairs_vs_nquasars}. Because the RM sample dominates our data in pair space and the non-RM sample dominates our data in unique quasar space, this figure also illustrates the differences in physical properties between the two samples. While the RM quasars more uniformly sample a wide range of masses and bolometric luminosities, the non-RM sample preferentially targets higher luminosity quasars with more massive black holes. This is likely an effect of RM targeting all quasars in a given field on sky above a flux limit, whereas the many observing strategies that influence the non-RM sample appear to target bright quasars overall in this redshift range.

\begin{figure}
\centering
\includegraphics[width=\linewidth]{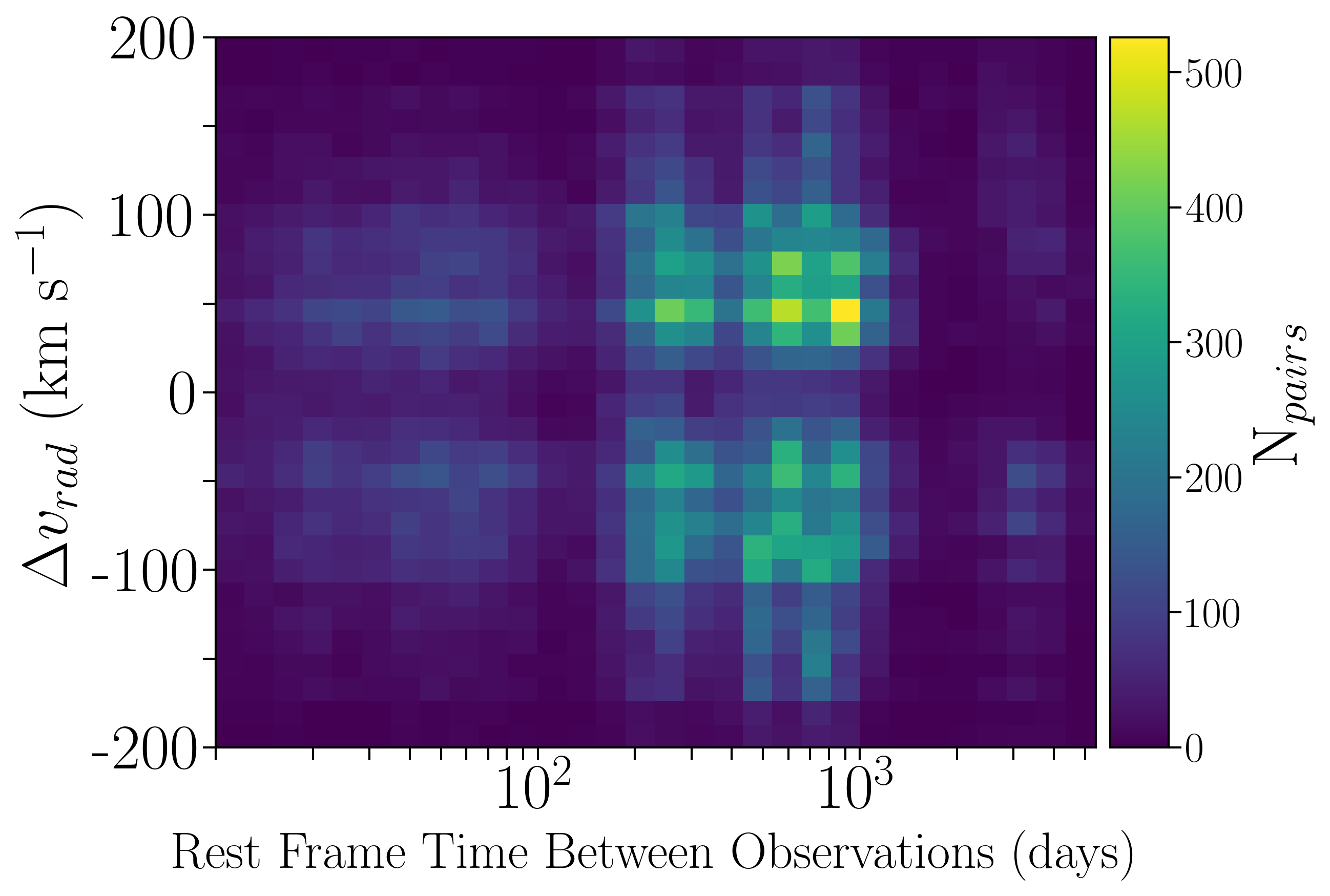}
\caption{2D histogram of change in radial velocity of the centroid of broad \Hb\ as a function of the rest-frame time between observations for each pair of spectra.}
\label{fig:jitter_dt}
\end{figure}

We show a two-dimensional histogram of the change in radial velocity of broad \Hb\ as a function of the rest-frame time between observations for pairs of spectra in our sample in Fig.~\ref{fig:jitter_dt}. Over-densities along vertical stripes are caused by the seasonal observing cadence of SDSS, with the most prominent over-density, between $\Delta t$ of 200 and 1,000 days mostly consisting of RM observations. Horizontal stripes in the data are caused by the 30 km s$^{-1}$ sampling of the cross-correlation in the $\Delta v_{rad}$ calculation. The uncertainty caused by this sampling is added in quadrature but is not the biggest source of uncertainty in $\Delta v_{rad}$. 

\begin{figure*}
\centering
\includegraphics[width=\linewidth]{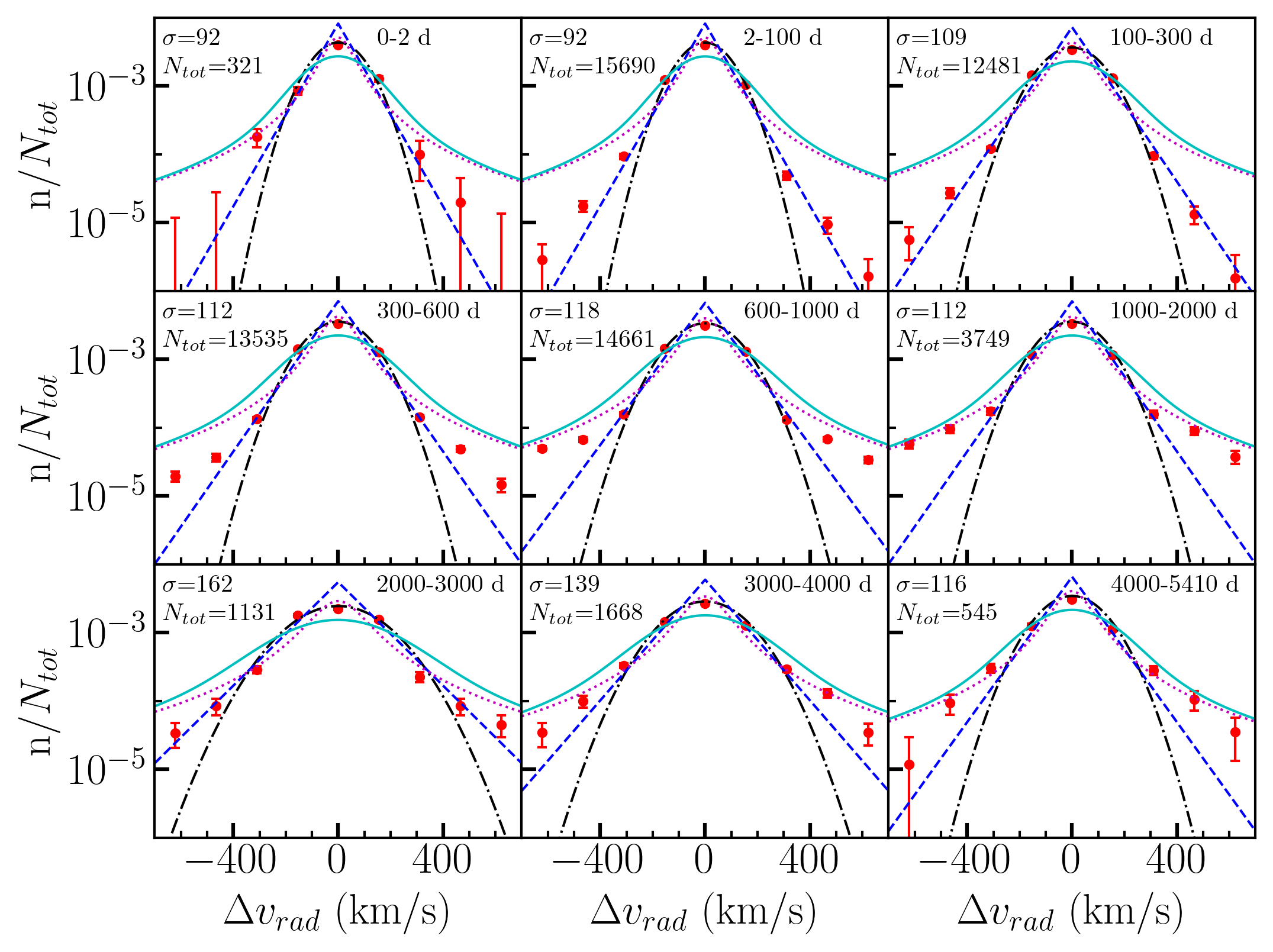}
\caption{Histograms for the change in centroid of the broad \Hb\ emission line for different ranges of rest frame time between observations. The range of rest frame times in days is printed in the top right of each panel. $N_{tot}$ is the total number of pairs used for each histogram, n is the number of pairs in a bin divided by the bin width, and $\sigma$=0.74(IQR) where IQR is the inter-quartile range for $\Delta v_{rad}$ values in each histogram. A Gaussian (black, dot-dash), exponential (blue, dash), Lorentzian (purple, dot), and Voigt (cyan, solid) distribution are plotted on each histogram with widths determined by the IQR of the data in each histogram (for normal distribution, $\sigma$=0.74 IQR, for Lorentz distribution, $\gamma$=0.5 IQR). The error bars in each bin are calculated with a Monte Carlo simulation (Section \ref{jitter}). In Figure Set 1, we show similar histograms for the change in 5100 \AA\ continuum luminosity, the change in luminosity of broad \Hb, and the change in FWHM of broad \Hb, as well as a recreation of all four plots excluding any RM sample pairs. The complete figure set (8 images) is available in the online journal.}
\label{fig:delta_vrad_hists}
\end{figure*}

\begin{figure*}
\centering
\includegraphics[width=\linewidth]{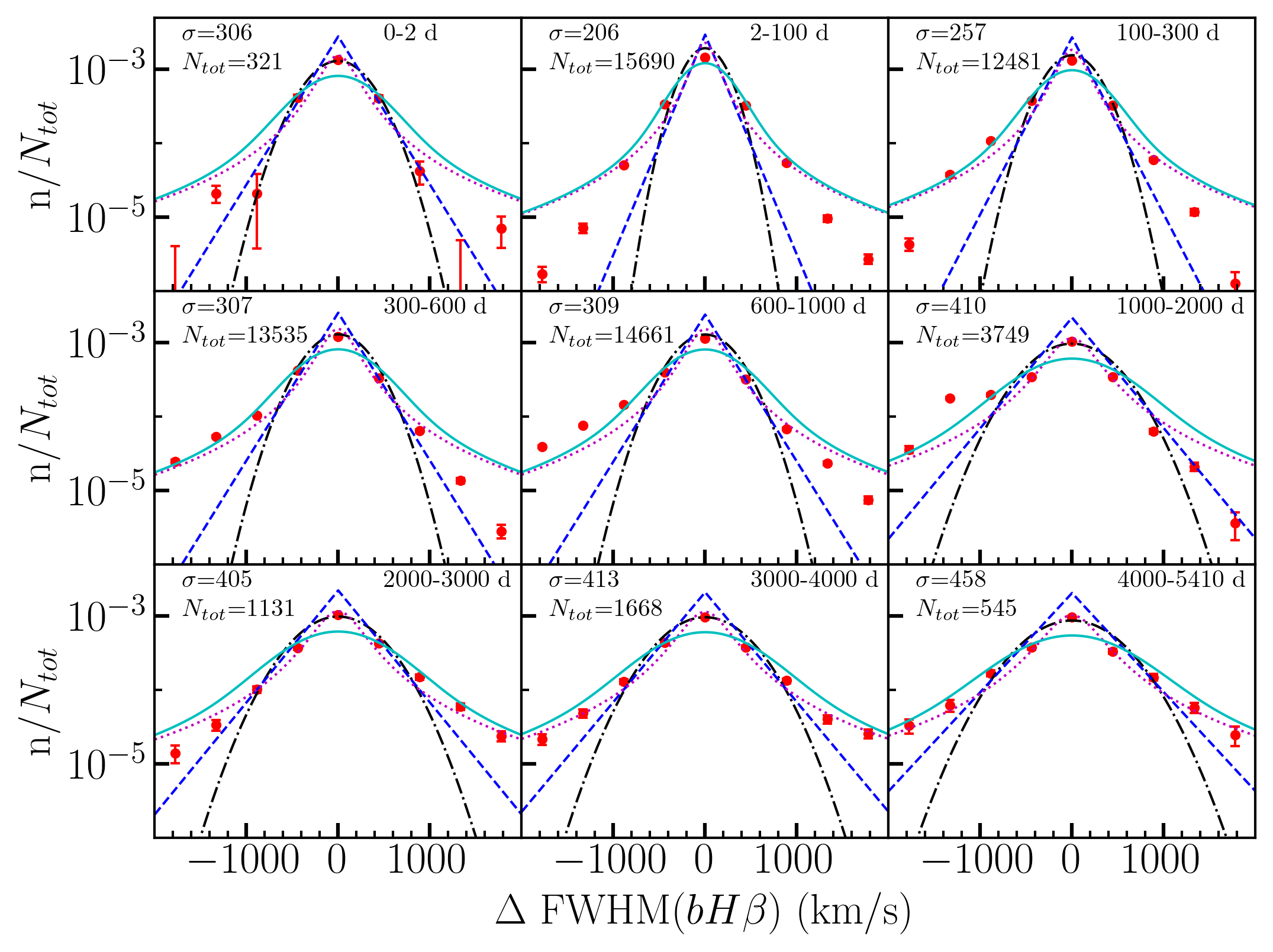}
\caption{Histograms for the change in FWHM of the broad \Hb\ emission line for different ranges of rest frame time between observations. All formatting is the same as in Fig. \ref{fig:delta_vrad_hists}.}
\label{fig:delta_fwhm_hists}
\end{figure*}

In Fig. Set 1, we plot histograms for four different broad \Hb\ variability metrics, $\Delta v_{rad}$, $\Delta$log(L$_{5100}$/erg s$^{-1}$), $\Delta$log(L$_{bH\beta}$/erg s$^{-1}$), and $\Delta$FWHM(bH$\beta$), for different ranges of rest frame time between observations. We also plot the same histograms for the non-RM sub-sample. Fig. \ref{fig:delta_vrad_hists} plots the histogram for $\Delta v_{rad}$, and Fig. \ref{fig:delta_fwhm_hists} plots the histogram for $\Delta$FWHM(bH$\beta$). We follow the method adopted in \cite{macleod2012}. Black, dot-dashed lines are Gaussian distributions with standard deviation of 0.74(IQR), where IQR is the inter-quartile range of the values in each histogram (the standard deviation of a normal distribution is equal to 0.74 times its IQR). Blue, dashed lines are exponential distributions with exp($-\Delta v_{rad}/\Delta_c$) where $\Delta_c=\sigma/\sqrt{2}$. Purple, dotted lines are Lorentz distributions with $\gamma=0.5$(IQR), where $\gamma$ is the half-width half maximum (HWHM; the HWHM of a Lorentz distribution is equal to half its IQR). Cyan, solid lines show a Voigt profile, which is calculated via both $\sigma$ and $\gamma$. The error bars in each bin are calculated with a Monte-Carlo simulation wherein all of the values in each histogram are resampled according to their uncertainties, creating changes in bin-membership. Repeating this resampling 100 times creates a distribution of the number of pairs in a given bin. 

The histogram covering baselines of less than two days (top left) can be used as a noise estimate, under the assumption that as $\Delta t$ approaches zero, the true variability approaches zero (Section \ref{lowdt}), with excess variance at longer baselines being indicative of intrinsic variability.

In Fig.~\ref{fig:delta_vrad_hists}, we show that $\Delta v_{rad}$ is not normally distributed. For rest-frame baselines of up to roughly 300 days, $\Delta v_{rad}$ is best represented by an exponential distribution, and for baselines greater than roughly 1,000 days the distribution is broadly consistent with a Lorentzian distribution. The width of the best-fit Lorentz distribution for baselines longer than 1,000 rest-frame days is $\gamma=81$ km s$^{-1}$. 

The fact that $\Delta v_{rad}$ is not normally distributed is at odds with the results of prior analyses. For example, \cite{doan2020} estimate changes in the radial velocity of the centroid of broad \Hb\ by fitting a third order polynomial to the radial velocity curves of the blue and red peaks of 3C 390.3, covering roughly 30 years of observations. They find that the residuals of the third order polynomial fit are roughly normally distributed. They conclude that the polynomial fit represents the long-term variability of the centroid, and the Gaussian residuals represent the short term variability. However, we observe that our short-baseline variability is not normally distributed. 

Fig.~\ref{fig:delta_fwhm_hists} shows a significant asymmetry in the change in FWHM of broad \Hb\ for baselines of roughly 300 to 2,000 days, such that a decrease in the FWHM over time is more likely than an increase. The asymmetry is almost completely suppressed when RM quasars are removed from the sample, suggesting that it may be dominated by a few RM quasars. 

It may be the case that a few RM quasars had an earlier SDSS observation when the quasar was in a relatively dim state with broader \Hb\ emission, followed by an RM observing campaign years later with many observations when the quasar was coincidentally in a brighter state with narrower \Hb\ emission. This would create many pairs of spectra per quasar with observational baselines of years, each with a negative $\Delta$FWHM(b\Hb). While the 300-600 day and the 600-1,000 day panels each have roughly 14,000 pairs of observations, enough to prevent individual RM quasars from dominating the entire distribution, the asymmetry occurs in the wing of the distribution, significantly more than an order of magnitude below the peak.

\begin{figure}
\centering
\includegraphics[width=\linewidth]{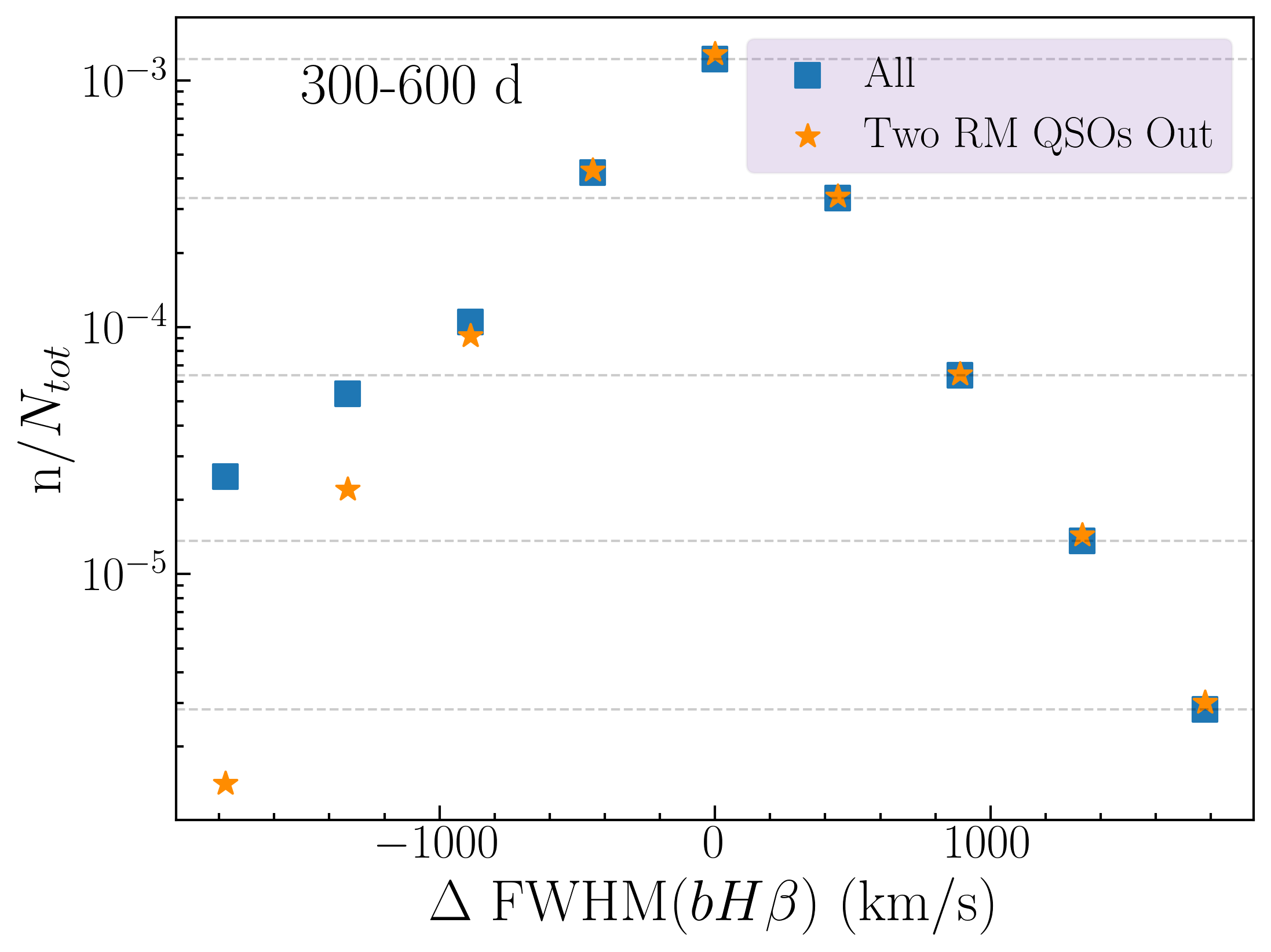}
\caption{Illustration of the effect of individual RM quasars on the observed asymmetry in the histograms for $\Delta$FWHM(b\Hb). Blue squares are a histogram for all pairs of observations (not just RM quasars) with rest-frame baselines between 300 and 600 days, created with the same method as in Fig. \ref{fig:delta_fwhm_hists}. Horizontal lines are plotted at the values of the blue squares with $\Delta$FWHM(b\Hb)$\geq$0 to illustrate where the related points on the left side of the plot would lie if the distribution were symmetric. Orange stars are a histogram of the same sample with all of the pairs of observations from two RM quasars removed. These two quasars were hand selected for their predominantly negative values for $\Delta$FWHM(b\Hb). Removing these two RM quasars almost completely removes the asymmetry in $\Delta$FWHM(b\Hb).}
\label{fig:dfwhm_rm}
\end{figure}

To test this effect, we hand selected two RM quasars for their predominantly negative values for $\Delta$FWHM(b\Hb) and removed them from the sample, and then plotted histograms for $\Delta$FWHM(b\Hb) for the full sample and for the sample without these two RM quasars. As shown in Fig. \ref{fig:dfwhm_rm}, removing these two RM quasars almost completely removes the observed asymmetry.

\figsetstart

\figsetnum{1}
\figsettitle{Variability Histograms}
\figsetgrpstart
\figsetgrpnum{figurenumber.1}
\figsetgrptitle{$\Delta v_{rad}$}
\figsetplot{delta_vrad_hists_mc.eps}
\figsetgrpnote{Histograms for the change in centroid of the broad \Hb\ emission line for different ranges of rest frame time between observations. The range of rest frame times in days is printed in the top right of each panel. $N_{tot}$ is the total number of pairs used for each histogram, n is the number of pairs in a bin divided by the bin width, and $\sigma$=0.74(IQR) where IQR is the inter-quartile range for $\Delta v_{rad}$ values in each histogram. A Gaussian (black, dot-dash), exponential (blue, dash), Lorentzian (purple, dot), and Voigt (cyan, solid) distribution are plotted on each histogram with widths determined by the IQR of the data in each histogram (for normal distribution, $\sigma$=0.74 IQR, for Lorentz distribution, $\gamma$=0.5 IQR). The error bars in each bin are calculated with a Monte Carlo simulation.}
\figsetgrpend

\figsetgrpstart
\figsetgrpnum{figurenumber.2}
\figsetgrptitle{$\Delta log L_{5100}$}
\figsetplot{delta_l5100_hists_mc.eps}
\figsetgrpnote{Histograms for the change in 5100 $\AA$ continuum luminosity for different ranges of rest frame time between observations. All formatting is the same as in Fig. 1}
\figsetgrpend

\figsetgrpstart
\figsetgrpnum{figurenumber.3}
\figsetgrptitle{$\Delta log L_{H\beta}$}
\figsetplot{delta_lHb_hists_mc.eps}
\figsetgrpnote{Histograms for the change in luminosity of broad \Hb\ for different ranges of rest frame time between observations. All formatting is the same as in Fig. 1}
\figsetgrpend

\figsetgrpstart
\figsetgrpnum{figurenumber.4}
\figsetgrptitle{$\Delta$ FWHM}
\figsetplot{delta_fwhm_hists_mc.eps}
\figsetgrpnote{Histograms for the change in FWHM of broad \Hb\ for different ranges of rest frame time between observations. All formatting is the same as in Fig. 1}
\figsetgrpend

\figsetgrpstart
\figsetgrpnum{figurenumber.5}
\figsetgrptitle{$\Delta v_{rad}$ No RM}
\figsetplot{delta_vrad_no_rm_hists_mc.eps}
\figsetgrpnote{Histograms for the change in centroid of the broad \Hb\ emission line for different ranges of rest frame time between observations, excluding any pairs in the RM sample. All formatting is the same as in Fig. 1}
\figsetgrpend

\figsetgrpstart
\figsetgrpnum{figurenumber.6}
\figsetgrptitle{$\Delta log L_{5100}$ No RM}
\figsetplot{delta_l5100_no_rm_hists_mc.eps}
\figsetgrpnote{Histograms for the change in 5100 \AA\ continuum luminosity for different ranges of rest frame time between observations, excluding any pairs in the RM sample. All formatting is the same as in Fig. 1}
\figsetgrpend

\figsetgrpstart
\figsetgrpnum{figurenumber.7}
\figsetgrptitle{$\Delta log L_{H\beta}$ No RM} 
\figsetplot{delta_lHb_no_rm_hists_mc.eps}
\figsetgrpnote{Histograms for the change in luminosity of broad \Hb\ for different ranges of rest frame time between observations, excluding any pairs in the RM sample. All formatting is the same as in Fig. 1}
\figsetgrpend

\figsetgrpstart
\figsetgrpnum{figurenumber.8}
\figsetgrptitle{$\Delta$ FWHM No RM}
\figsetplot{delta_fwhm_no_rm_hists_mc.eps}
\figsetgrpnote{Histograms for the change in FWHM of broad \Hb\ for different ranges of rest frame time between observations, excluding any pairs in the RM sample. All formatting is the same as in Fig. 1}
\figsetgrpend

\figsetend

\section{Discussion}\label{discussion}

In this work we have presented the methodology and first measurements that form the basis for further time-domain analysis with SDSS, especially for future work with SDSS~V. Any large-sample spectroscopic variability studies with SDSS have to contend with the effects of the difference in cadence between the RM sample and the rest of the quasars that have been re-observed. While RM quasars make up a small fraction of the quasars with repeat spectra in SDSS, they make up the vast majority of the pairs of spectra (Fig. \ref{fig:dt_hist_rm}). RM quasars have distinctly different physical properties than non-RM quasars, most notably more evenly sampling a wider range of masses than the non-RM sample, which preferentially selects high mass quasars, even at low redshift (Fig. \ref{fig:npairs_vs_nquasars}). 

The RM sample also has a different observing cadence than the non-RM sample. Baselines from roughly $\Delta t$ of 200 to 1000 days are extremely well-sampled with RM observations, while baselines greater than roughly 2000 days are dominated by the non-RM sample (Fig. \ref{fig:dt_hist_rm}). Thus, any variability analysis must carefully consider its binning in $\Delta t$, as different binnings can represent substantially different underlying populations.

The measurements we have made, especially of the radial velocity changes in the broad \Hb\ line profile, will have important implications in the context of understanding the central engines of quasars. We still lack a predictive physical model for the BLR in normal quasars. Such predictive models may need different physical prescriptions for short timescale variability, which corresponds with the light-crossing time of the BLR, and long timescale variability, which corresponds with the dynamical time of the BLR. As such, having observational constraints on the variability that cover these timescales will serve as a crucial benchmark for testing future theoretical models.

Our measurements provide observational constraints on population-level quasar profile variability that are necessary to inform searches for supermassive black hole binaries (SMBHBs). Spectroscopic searches for SMBHBs aim to find sinusoidal Doppler shifts in the centroid of a broad emission line caused by the orbital motion of a single black hole in the binary. Because the expected orbital periods for SMBHBs can be many decades, studies will instead search for monotonic changes in the centroid of a broad emission line over many years, corresponding to a portion of a single binary orbit \citep{eracleous2012,shen2013,runnoe2015,runnoe2017,guo2019}. These studies must contend with the confusion factor of centroid changes for normal, single black hole quasars. Having an empirical constraint on population level centroid variability will better inform these works, allowing them to separate signatures of SMBHBs from those of normal quasars. Additionally, prior SMBHB studies have assumed that centroid changes on short-timescales in normal quasars have Gaussian distributions, which we have shown to be untrue for this sample \citep{guo2019, doan2020}. 

\section{Summary \& Conclusions}\label{summary}

We have created a catalog of spectroscopic variability properties for all $z<0.8$ SDSS quasars with publicly available data through DR16Q \citep{lyke2020}. Our sample consists of 10,336 quasars and 31,320 spectra for those quasars with baselines between observations of up to 5410 days in the rest frame, including many thousands of pairs of spectra with baselines longer than a year. We performed a spectral decomposition of every spectrum in our sample, allowing us to subtract out all components of the spectrum except for broad \Hb\ emission. We then measured properties of the broad \Hb\ line profiles and their changes, including the change in radial velocity of the emission line ($\Delta v_{rad}$), as well as the luminosity, FWHM, equivalent width and Pearson skewness coefficient, with a thorough estimate of their uncertainties. 

We have provided a table with measured properties for each spectrum, including an estimate for the mass and the bolometric luminosity, and a table with the difference in measured properties for each pair of spectra, including $\Delta v_{rad}$. We find black hole masses and bolometric luminosities which are broadly consistent with other measurements of SDSS quasar spectra for a low redshift sample \citep{wushen2022}.

We discuss two methods for estimating the uncertainty in $\Delta v_{rad}$, in the case where a spectral decomposition is used to isolate the broad line. The first method involves fitting a pilot sample with MCMC for its robust uncertainty estimates and taking samples from the posterior distribution for both of the spectra in the pair to build up a distribution of $\Delta v_{rad}$ estimates. One can then fit a scaling relation for the uncertainty in $\Delta v_{rad}$ as a function of relevant independent variable(s), in our case the $\Delta v_{rad}$ itself and the SNR of the broad line, SNR(b\Hb), to create an uncertainty estimate that can be applied to all available pairs of spectra. This method has the advantage of accounting for both the uncertainty in the decomposition as well as the uncertainty in the cross-correlation measurement of $\Delta v_{rad}$. It is also able to produce an uncertainty estimate for individual pairs of spectra. But it is highly dependent on the method chosen for fitting the scaling relation. The second method involves taking pairs of spectra with baselines short enough that the true variability in the pair can be assumed to be zero, and taking the observed variability as a sample from the noise distribution. This method has the advantage of being non-parametric, though it does inherit the model choices made in the spectral decomposition. However, it is unable to provide uncertainty estimates for individual pairs of spectra and can only be used as an ensemble estimate.

We find that, in general, $\Delta v_{rad}$ is not normally distributed. For shorter rest frame times between observations of up to roughly 300 days, it is best represented by an exponential distribution. For longer baselines, the distribution appears Lorentzian. We provide the widths of the $\Delta v_{rad}$ distributions, measured through the IQR, for different ranges of rest frame $\Delta$t.

We discuss the effect of the difference in cadence between the RM sample and the rest of the quasars in SDSS that are observed at least twice. The RM subsample accounts for a very small fraction of the quasars in our sample but the vast majority of the pairs of spectra. This difference is important because the RM quasars have a different selection function, and therefore different underlying physical properties, from other quasars in SDSS (Fig. \ref{fig:npairs_vs_nquasars}). It can also create interesting small number statistics effects where the behavior of individual RM quasar(s) can appreciably affect the observed behavior of the entire sample. We show that an observed asymmetry in the FWHM of broad \Hb\ can be almost completely suppressed by removing two RM quasars from the sample (Fig. \ref{fig:dfwhm_rm}). The methods and data from this study will inform future work on spectroscopic variability with SDSS.

\section*{Acknowledgments}

C.M.D and J.C.R. acknowledge support from the National Science Foundation (NSF) from grant NSF AST-2205719 and the NASA Preparatory Science program under award 20-LPS20-0013.

M.E. acknowledges support by the National Science Foundation through grant AST-2205720 (WoU-MMA).

Funding for the Sloan Digital Sky Survey IV was provided by the Alfred P. Sloan Foundation, the U.S. Department of Energy Office of Science, and the Participating Institutions. SDSS~IV acknowledges support and resources from the Center for High Performance Computing at the University of Utah. The SDSS website is \url{http://www.sdss.org/}.

SDSS~IV is managed by the Astrophysical Research Consortium for the Participating Institutions of the SDSS Collaboration including the Brazilian Participation Group, the Carnegie Institution for Science, Carnegie Mellon University, the Chilean Participation Group, the French Participation Group, Harvard-Smithsonian Center for Astrophysics, Instituto de Astrofisica de Canarias, The Johns Hopkins University, Kavli Institute for the Physics and Mathematics of the Universe (IPMU)/University of Tokyo, Lawrence Berkeley National Laboratory, Leibniz Institut fur Astrophysik Potsdam (AIP), Max-Planck-Institut fur Astronomie (MPIA Heidelberg), Max-Planck-Institut fur Astrophysik (MPA Garching), Max-Planck-Institut fur Extraterrestrische Physik (MPE), National Astronomical Observatories of China, New Mexico State University, New York University, University of Notre Dame, Observatario  Nacional/MCTI, The Ohio State University, Pennsylvania State University, Shanghai Astronomical Observatory, United Kingdom Participation Group, Universidad Nacional Autonoma de Mexico, University of Arizona, University of Colorado Boulder, University of Oxford, University of Portsmouth, University of Utah, University of Virginia, University of Washington, University of Wisconsin, Vanderbilt University, and Yale University.

\bibliography{cites.bib}{}

\begin{thebibliography}{}
\expandafter\ifx\csname natexlab\endcsname\relax\def\natexlab#1{#1}\fi
\providecommand{\url}[1]{\href{#1}{#1}}
\providecommand{\dodoi}[1]{doi:~\href{http://doi.org/#1}{\nolinkurl{#1}}}
\providecommand{\doeprint}[1]{\href{http://ascl.net/#1}{\nolinkurl{http://ascl.net/#1}}}
\providecommand{\doarXiv}[1]{\href{https://arxiv.org/abs/#1}{\nolinkurl{https://arxiv.org/abs/#1}}}

\bibitem[{N. {Arav} {et~al.}(1997){Arav}, {Barlow}, {Laor}, \& {Blandford}}]{arav1997}
{Arav}, N., {Barlow}, T.~A., {Laor}, A., \& {Blandford}, R.~D. 1997, \bibinfo{title}{{Keck high-resolution spectroscopy of MRK 335: constraints on the number of emitting clouds in the broad-line region},} \mnras, 288, 1015, \dodoi{10.1093/mnras/288.4.1015}

\bibitem[{N. {Arav} {et~al.}(1998){Arav}, {Barlow}, {Laor}, {Sargent}, \& {Blandford}}]{arav1998}
{Arav}, N., {Barlow}, T.~A., {Laor}, A., {Sargent}, W. L.~W., \& {Blandford}, R.~D. 1998, \bibinfo{title}{{Are AGN broad emission lines formed by discrete clouds? Analysis of Keck high-resolution spectroscopy of NGC 4151},} \mnras, 297, 990, \dodoi{10.1046/j.1365-8711.1998.297004990.x}

\bibitem[{M. {Batiste} {et~al.}(2017){Batiste}, {Bentz}, {Raimundo}, {Vestergaard}, \& {Onken}}]{batiste2017}
{Batiste}, M., {Bentz}, M.~C., {Raimundo}, S.~I., {Vestergaard}, M., \& {Onken}, C.~A. 2017, \bibinfo{title}{{Recalibration of the M $_{BH}$-{\ensuremath{\sigma}} $_{{\ensuremath{\star}}}$ Relation for AGN},} \apjl, 838, L10, \dodoi{10.3847/2041-8213/aa6571}

\bibitem[{M.~C. {Bentz} {et~al.}(2009){Bentz}, {Walsh}, {Barth}, {Baliber}, {Bennert}, {Canalizo}, {Filippenko}, {Ganeshalingam}, {Gates}, {Greene}, {Hidas}, {Hiner}, {Lee}, {Li}, {Malkan}, {Minezaki}, {Sakata}, {Serduke}, {Silverman}, {Steele}, {Stern}, {Street}, {Thornton}, {Treu}, {Wang}, {Woo}, \& {Yoshii}}]{bentz2009}
{Bentz}, M.~C., {Walsh}, J.~L., {Barth}, A.~J., {et~al.} 2009, \bibinfo{title}{{The Lick AGN Monitoring Project: Broad-line Region Radii and Black Hole Masses from Reverberation Mapping of H{\ensuremath{\beta}}},} \apj, 705, 199, \dodoi{10.1088/0004-637X/705/1/199}

\bibitem[{R.~D. {Blandford} \& C.~F. {McKee}(1982){Blandford} \& {McKee}}]{blandfordmckee}
{Blandford}, R.~D., \& {McKee}, C.~F. 1982, \bibinfo{title}{{Reverberation mapping of the emission line regions of Seyfert galaxies and quasars.},} \apj, 255, 419, \dodoi{10.1086/159843}

\bibitem[{M.~R. {Blanton} {et~al.}(2017){Blanton}, {Bershady}, {Abolfathi}, {Albareti}, {Allende Prieto}, {Almeida}, {Alonso-Garc{\'\i}a}, {Anders}, {Anderson}, {Andrews}, {Aquino-Ort{\'\i}z}, {Arag{\'o}n-Salamanca}, {Argudo-Fern{\'a}ndez}, {Armengaud}, {Aubourg}, {Avila-Reese}, {Badenes}, {Bailey}, {Barger}, {Barrera-Ballesteros}, {Bartosz}, {Bates}, {Baumgarten}, {Bautista}, {Beaton}, {Beers}, {Belfiore}, {Bender}, {Berlind}, {Bernardi}, {Beutler}, {Bird}, {Bizyaev}, {Blanc}, {Blomqvist}, {Bolton}, {Boquien}, {Borissova}, {van den Bosch}, {Bovy}, {Brandt}, {Brinkmann}, {Brownstein}, {Bundy}, {Burgasser}, {Burtin}, {Busca}, {Cappellari}, {Delgado Carigi}, {Carlberg}, {Carnero Rosell}, {Carrera}, {Chanover}, {Cherinka}, {Cheung}, {G{\'o}mez Maqueo Chew}, {Chiappini}, {Choi}, {Chojnowski}, {Chuang}, {Chung}, {Cirolini}, {Clerc}, {Cohen}, {Comparat}, {da Costa}, {Cousinou}, {Covey}, {Crane}, {Croft}, {Cruz-Gonzalez}, {Garrido Cuadra}, {Cunha}, {Damke}, {Darling}, {Davies}, {Dawson}, {de la Macorra},
  {Dell'Agli}, {De Lee}, {Delubac}, {Di Mille}, {Diamond-Stanic}, {Cano-D{\'\i}az}, {Donor}, {Downes}, {Drory}, {du Mas des Bourboux}, {Duckworth}, {Dwelly}, {Dyer}, {Ebelke}, {Eigenbrot}, {Eisenstein}, {Emsellem}, {Eracleous}, {Escoffier}, {Evans}, {Fan}, {Fern{\'a}ndez-Alvar}, {Fernandez-Trincado}, {Feuillet}, {Finoguenov}, {Fleming}, {Font-Ribera}, {Fredrickson}, {Freischlad}, {Frinchaboy}, {Fuentes}, {Galbany}, {Garcia-Dias}, {Garc{\'\i}a-Hern{\'a}ndez}, {Gaulme}, {Geisler}, {Gelfand}, {Gil-Mar{\'\i}n}, {Gillespie}, {Goddard}, {Gonzalez-Perez}, {Grabowski}, {Green}, {Grier}, {Gunn}, {Guo}, {Guy}, {Hagen}, {Hahn}, {Hall}, {Harding}, {Hasselquist}, {Hawley}, {Hearty}, {Gonzalez Hern{\'a}ndez}, {Ho}, {Hogg}, {Holley-Bockelmann}, {Holtzman}, {Holzer}, {Huehnerhoff}, {Hutchinson}, {Hwang}, {Ibarra-Medel}, {da Silva Ilha}, {Ivans}, {Ivory}, {Jackson}, {Jensen}, {Johnson}, {Jones}, {J{\"o}nsson}, {Jullo}, {Kamble}, {Kinemuchi}, {Kirkby}, {Kitaura}, {Klaene}, {Knapp}, {Kneib}, {Kollmeier}, {Lacerna}, {Lane},
  {Lang}, {Law}, {Lazarz}, {Lee}, {Le Goff}, {Liang}, {Li}, {Li}, {Lian}, {Lima}, {Lin}, {Lin}, {Bertran de Lis}, {Liu}, {de Icaza Lizaola}, {Long}, {Lucatello}, {Lundgren}, {MacDonald}, {Deconto Machado}, {MacLeod}, {Mahadevan}, {Geimba Maia}, {Maiolino}, {Majewski}, {Malanushenko}, {Malanushenko}, {Manchado}, {Mao}, {Maraston}, {Marques-Chaves}, {Masseron}, {Masters}, {McBride}, {McDermid}, {McGrath}, {McGreer}, {Medina Pe{\~n}a}, {Melendez}, {Merloni}, {Merrifield}, {Meszaros}, {Meza}, {Minchev}, {Minniti}, {Miyaji}, {More}, {Mulchaey}, {M{\"u}ller-S{\'a}nchez}, {Muna}, {Munoz}, {Myers}, {Nair}, {Nandra}, {Correa do Nascimento}, {Negrete}, {Ness}, {Newman}, {Nichol}, {Nidever}, {Nitschelm}, {Ntelis}, {O'Connell}, {Oelkers}, {Oravetz}, {Oravetz}, {Pace}, {Padilla}, {Palanque-Delabrouille}, {Alonso Palicio}, {Pan}, {Parejko}, {Parikh}, {P{\^a}ris}, {Park}, {Patten}, {Peirani}, {Pellejero-Ibanez}, {Penny}, {Percival}, {Perez-Fournon}, {Petitjean}, {Pieri}, {Pinsonneault}, {Pisani}, {Poleski}, {Prada},
  {Prakash}, {Queiroz}, {Raddick}, {Raichoor}, {Barboza Rembold}, {Richstein}, {Riffel}, {Riffel}, {Rix}, {Robin}, {Rockosi}, {Rodr{\'\i}guez-Torres}, {Roman-Lopes}, {Rom{\'a}n-Z{\'u}{\~n}iga}, {Rosado}, {Ross}, {Rossi}, {Ruan}, {Ruggeri}, {Rykoff}, {Salazar-Albornoz}, {Salvato}, {S{\'a}nchez}, {Aguado}, {S{\'a}nchez-Gallego}, {Santana}, {Santiago}, {Sayres}, {Schiavon}, {da Silva Schimoia}, {Schlafly}, {Schlegel}, {Schneider}, {Schultheis}, {Schuster}, {Schwope}, {Seo}, {Shao}, {Shen}, {Shetrone}, {Shull}, {Simon}, {Skinner}, {Skrutskie}, {Slosar}, {Smith}, {Sobeck}, {Sobreira}, {Somers}, {Souto}, {Stark}, {Stassun}, {Stauffer}, {Steinmetz}, {Storchi-Bergmann}, {Streblyanska}, {Stringfellow}, {Su{\'a}rez}, {Sun}, {Suzuki}, {Szigeti}, {Taghizadeh-Popp}, {Tang}, {Tao}, {Tayar}, {Tembe}, {Teske}, {Thakar}, {Thomas}, {Thompson}, {Tinker}, {Tissera}, {Tojeiro}, {Hernandez Toledo}, {de la Torre}, {Tremonti}, {Troup}, {Valenzuela}, {Martinez Valpuesta}, {Vargas-Gonz{\'a}lez}, {Vargas-Maga{\~n}a}, {Vazquez},
  {Villanova}, {Vivek}, {Vogt}, {Wake}, {Walterbos}, {Wang}, {Weaver}, {Weijmans}, {Weinberg}, {Westfall}, {Whelan}, {Wild}, {Wilson}, {Wood-Vasey}, {Wylezalek}, {Xiao}, {Yan}, {Yang}, {Ybarra}, {Y{\`e}che}, {Zakamska}, {Zamora}, {Zarrouk}, {Zasowski}, {Zhang}, {Zhao}, {Zheng}, {Zheng}, {Zhou}, {Zhou}, {Zhu}, {Zoccali}, \& {Zou}}]{blanton2017}
{Blanton}, M.~R., {Bershady}, M.~A., {Abolfathi}, B., {et~al.} 2017, \bibinfo{title}{{Sloan Digital Sky Survey IV: Mapping the Milky Way, Nearby Galaxies, and the Distant Universe},} \aj, 154, 28, \dodoi{10.3847/1538-3881/aa7567}

\bibitem[{A.~S. {Bolton} {et~al.}(2012){Bolton}, {Schlegel}, {Aubourg}, {Bailey}, {Bhardwaj}, {Brownstein}, {Burles}, {Chen}, {Dawson}, {Eisenstein}, {Gunn}, {Knapp}, {Loomis}, {Lupton}, {Maraston}, {Muna}, {Myers}, {Olmstead}, {Padmanabhan}, {P{\^a}ris}, {Percival}, {Petitjean}, {Rockosi}, {Ross}, {Schneider}, {Shu}, {Strauss}, {Thomas}, {Tremonti}, {Wake}, {Weaver}, \& {Wood-Vasey}}]{bolton2012}
{Bolton}, A.~S., {Schlegel}, D.~J., {Aubourg}, {\'E}., {et~al.} 2012, \bibinfo{title}{{Spectral Classification and Redshift Measurement for the SDSS-III Baryon Oscillation Spectroscopic Survey},} \aj, 144, 144, \dodoi{10.1088/0004-6256/144/5/144}

\bibitem[{T.~A. {Boroson} \& R.~F. {Green}(1992){Boroson} \& {Green}}]{boroson&green}
{Boroson}, T.~A., \& {Green}, R.~F. 1992, \bibinfo{title}{{The Emission-Line Properties of Low-Redshift Quasi-stellar Objects},} \apjs, 80, 109, \dodoi{10.1086/191661}

\bibitem[{J.~A. {Cardelli} {et~al.}(1989){Cardelli}, {Clayton}, \& {Mathis}}]{ccm}
{Cardelli}, J.~A., {Clayton}, G.~C., \& {Mathis}, J.~S. 1989, \bibinfo{title}{{The relationship between IR, optical, and UV extinction.},} in Interstellar Dust, ed. L.~J. {Allamandola} \& A.~G.~G.~M. {Tielens}, Vol. 135, 5--10

\bibitem[{S.~M. {Croom} {et~al.}(2001){Croom}, {Smith}, {Boyle}, {Shanks}, {Loaring}, {Miller}, \& {Lewis}}]{croom2001}
{Croom}, S.~M., {Smith}, R.~J., {Boyle}, B.~J., {et~al.} 2001, \bibinfo{title}{{The 2dF QSO Redshift Survey - V. The 10k catalogue},} \mnras, 322, L29, \dodoi{10.1046/j.1365-8711.2001.04474.x}

\bibitem[{E. {Dalla Bont{\`a}} {et~al.}(2020){Dalla Bont{\`a}}, {Peterson}, {Bentz}, {Brandt}, {Ciroi}, {De Rosa}, {Fonseca Alvarez}, {Grier}, {Hall}, {Hern{\'a}ndez Santisteban}, {Ho}, {Homayouni}, {Horne}, {Kochanek}, {Li}, {Morelli}, {Pizzella}, {Pogge}, {Schneider}, {Shen}, {Trump}, \& {Vestergaard}}]{bonta2020}
{Dalla Bont{\`a}}, E., {Peterson}, B.~M., {Bentz}, M.~C., {et~al.} 2020, \bibinfo{title}{{The Sloan Digital Sky Survey Reverberation Mapping Project: Estimating Masses of Black Holes in Quasars with Single-epoch Spectroscopy},} \apj, 903, 112, \dodoi{10.3847/1538-4357/abbc1c}

\bibitem[{M. {Dietrich} {et~al.}(1999){Dietrich}, {Wagner}, {Courvoisier}, {Bock}, \& {North}}]{dietrich1998}
{Dietrich}, M., {Wagner}, S.~J., {Courvoisier}, T.~J.~L., {Bock}, H., \& {North}, P. 1999, \bibinfo{title}{{Structure of the broad-line region of 3C 273},} \aap, 351, 31

\bibitem[{A. {Doan} {et~al.}(2020){Doan}, {Eracleous}, {Runnoe}, {Liu}, {Mathes}, \& {Flohic}}]{doan2020}
{Doan}, A., {Eracleous}, M., {Runnoe}, J.~C., {et~al.} 2020, \bibinfo{title}{{An improved test of the binary black hole hypothesis for quasars with double-peaked broad Balmer lines},} \mnras, 491, 1104, \dodoi{10.1093/mnras/stz2705}

\bibitem[{A.~M. {Dumont} \& S. {Collin-Souffrin}(1990){Dumont} \& {Collin-Souffrin}}]{Dumont1990IV}
{Dumont}, A.~M., \& {Collin-Souffrin}, S. 1990, \bibinfo{title}{{Line and continuum emission from the outer regions of accretion discsin active galactic nuclei. IV. Line emission.},} \aap, 229, 313

\bibitem[{D.~J. {Eisenstein} {et~al.}(2011){Eisenstein}, {Weinberg}, {Agol}, {Aihara}, {Allende Prieto}, {Anderson}, {Arns}, {Aubourg}, {Bailey}, {Balbinot}, {Barkhouser}, {Beers}, {Berlind}, {Bickerton}, {Bizyaev}, {Blanton}, {Bochanski}, {Bolton}, {Bosman}, {Bovy}, {Brandt}, {Breslauer}, {Brewington}, {Brinkmann}, {Brown}, {Brownstein}, {Burger}, {Busca}, {Campbell}, {Cargile}, {Carithers}, {Carlberg}, {Carr}, {Chang}, {Chen}, {Chiappini}, {Comparat}, {Connolly}, {Cortes}, {Croft}, {Cunha}, {da Costa}, {Davenport}, {Dawson}, {De Lee}, {Porto de Mello}, {de Simoni}, {Dean}, {Dhital}, {Ealet}, {Ebelke}, {Edmondson}, {Eiting}, {Escoffier}, {Esposito}, {Evans}, {Fan}, {Femen{\'\i}a Castell{\'a}}, {Dutra Ferreira}, {Fitzgerald}, {Fleming}, {Font-Ribera}, {Ford}, {Frinchaboy}, {Garc{\'\i}a P{\'e}rez}, {Gaudi}, {Ge}, {Ghezzi}, {Gillespie}, {Gilmore}, {Girardi}, {Gott}, {Gould}, {Grebel}, {Gunn}, {Hamilton}, {Harding}, {Harris}, {Hawley}, {Hearty}, {Hennawi}, {Gonz{\'a}lez Hern{\'a}ndez}, {Ho}, {Hogg}, {Holtzman},
  {Honscheid}, {Inada}, {Ivans}, {Jiang}, {Jiang}, {Johnson}, {Jordan}, {Jordan}, {Kauffmann}, {Kazin}, {Kirkby}, {Klaene}, {Knapp}, {Kneib}, {Kochanek}, {Koesterke}, {Kollmeier}, {Kron}, {Lampeitl}, {Lang}, {Lawler}, {Le Goff}, {Lee}, {Lee}, {Leisenring}, {Lin}, {Liu}, {Long}, {Loomis}, {Lucatello}, {Lundgren}, {Lupton}, {Ma}, {Ma}, {MacDonald}, {Mack}, {Mahadevan}, {Maia}, {Majewski}, {Makler}, {Malanushenko}, {Malanushenko}, {Mandelbaum}, {Maraston}, {Margala}, {Maseman}, {Masters}, {McBride}, {McDonald}, {McGreer}, {McMahon}, {Mena Requejo}, {M{\'e}nard}, {Miralda-Escud{\'e}}, {Morrison}, {Mullally}, {Muna}, {Murayama}, {Myers}, {Naugle}, {Neto}, {Nguyen}, {Nichol}, {Nidever}, {O'Connell}, {Ogando}, {Olmstead}, {Oravetz}, {Padmanabhan}, {Paegert}, {Palanque-Delabrouille}, {Pan}, {Pandey}, {Parejko}, {P{\^a}ris}, {Pellegrini}, {Pepper}, {Percival}, {Petitjean}, {Pfaffenberger}, {Pforr}, {Phleps}, {Pichon}, {Pieri}, {Prada}, {Price-Whelan}, {Raddick}, {Ramos}, {Reid}, {Reyle}, {Rich}, {Richards}, {Rieke},
  {Rieke}, {Rix}, {Robin}, {Rocha-Pinto}, {Rockosi}, {Roe}, {Rollinde}, {Ross}, {Ross}, {Rossetto}, {S{\'a}nchez}, {Santiago}, {Sayres}, {Schiavon}, {Schlegel}, {Schlesinger}, {Schmidt}, {Schneider}, {Sellgren}, {Shelden}, {Sheldon}, {Shetrone}, {Shu}, {Silverman}, {Simmerer}, {Simmons}, {Sivarani}, {Skrutskie}, {Slosar}, {Smee}, {Smith}, {Snedden}, {Stassun}, {Steele}, {Steinmetz}, {Stockett}, {Stollberg}, {Strauss}, {Szalay}, {Tanaka}, {Thakar}, {Thomas}, {Tinker}, {Tofflemire}, {Tojeiro}, {Tremonti}, {Vargas Maga{\~n}a}, {Verde}, {Vogt}, {Wake}, {Wan}, {Wang}, {Weaver}, {White}, {White}, {Wilson}, {Wisniewski}, {Wood-Vasey}, {Yanny}, {Yasuda}, {Y{\`e}che}, {York}, {Young}, {Zasowski}, {Zehavi}, \& {Zhao}}]{eisenstein2011}
{Eisenstein}, D.~J., {Weinberg}, D.~H., {Agol}, E., {et~al.} 2011, \bibinfo{title}{{SDSS-III: Massive Spectroscopic Surveys of the Distant Universe, the Milky Way, and Extra-Solar Planetary Systems},} \aj, 142, 72, \dodoi{10.1088/0004-6256/142/3/72}

\bibitem[{M. {Elitzur} {et~al.}(2014){Elitzur}, {Ho}, \& {Trump}}]{elitzur2014}
{Elitzur}, M., {Ho}, L.~C., \& {Trump}, J.~R. 2014, \bibinfo{title}{{Evolution of broad-line emission from active galactic nuclei},} \mnras, 438, 3340, \dodoi{10.1093/mnras/stt2445}

\bibitem[{M. {Elitzur} \& H. {Netzer}(2016){Elitzur} \& {Netzer}}]{elitzur2016}
{Elitzur}, M., \& {Netzer}, H. 2016, \bibinfo{title}{{Disc outflows and high-luminosity true type 2 AGN},} \mnras, 459, 585, \dodoi{10.1093/mnras/stw657}

\bibitem[{R.~T. {Emmering} {et~al.}(1992){Emmering}, {Blandford}, \& {Shlosman}}]{Emmering1992}
{Emmering}, R.~T., {Blandford}, R.~D., \& {Shlosman}, I. 1992, \bibinfo{title}{{Magnetic Acceleration of Broad Emission-Line Clouds in Active Galactic Nuclei},} \apj, 385, 460, \dodoi{10.1086/170955}

\bibitem[{M. {Eracleous} {et~al.}(2012){Eracleous}, {Boroson}, {Halpern}, \& {Liu}}]{eracleous2012}
{Eracleous}, M., {Boroson}, T.~A., {Halpern}, J.~P., \& {Liu}, J. 2012, \bibinfo{title}{{A Large Systematic Search for Close Supermassive Binary and Rapidly Recoiling Black Holes},} \apjs, 201, 23, \dodoi{10.1088/0067-0049/201/2/23}

\bibitem[{ {Event Horizon Telescope Collaboration} {et~al.}(2019){Event Horizon Telescope Collaboration}, {Akiyama}, {Alberdi}, {Alef}, {Asada}, {Azulay}, {Baczko}, {Ball}, {Balokovi{\'c}}, {Barrett}, {Bintley}, {Blackburn}, {Boland}, {Bouman}, {Bower}, {Bremer}, {Brinkerink}, {Brissenden}, {Britzen}, {Broderick}, {Broguiere}, {Bronzwaer}, {Byun}, {Carlstrom}, {Chael}, {Chan}, {Chatterjee}, {Chatterjee}, {Chen}, {Chen}, {Cho}, {Christian}, {Conway}, {Cordes}, {Crew}, {Cui}, {Davelaar}, {De Laurentis}, {Deane}, {Dempsey}, {Desvignes}, {Dexter}, {Doeleman}, {Eatough}, {Falcke}, {Fish}, {Fomalont}, {Fraga-Encinas}, {Freeman}, {Friberg}, {Fromm}, {G{\'o}mez}, {Galison}, {Gammie}, {Garc{\'\i}a}, {Gentaz}, {Georgiev}, {Goddi}, {Gold}, {Gu}, {Gurwell}, {Hada}, {Hecht}, {Hesper}, {Ho}, {Ho}, {Honma}, {Huang}, {Huang}, {Hughes}, {Ikeda}, {Inoue}, {Issaoun}, {James}, {Jannuzi}, {Janssen}, {Jeter}, {Jiang}, {Johnson}, {Jorstad}, {Jung}, {Karami}, {Karuppusamy}, {Kawashima}, {Keating}, {Kettenis}, {Kim}, {Kim}, {Kim},
  {Kino}, {Koay}, {Koch}, {Koyama}, {Kramer}, {Kramer}, {Krichbaum}, {Kuo}, {Lauer}, {Lee}, {Li}, {Li}, {Lindqvist}, {Liu}, {Liuzzo}, {Lo}, {Lobanov}, {Loinard}, {Lonsdale}, {Lu}, {MacDonald}, {Mao}, {Markoff}, {Marrone}, {Marscher}, {Mart{\'\i}-Vidal}, {Matsushita}, {Matthews}, {Medeiros}, {Menten}, {Mizuno}, {Mizuno}, {Moran}, {Moriyama}, {Moscibrodzka}, {M{\"u}ller}, {Nagai}, {Nagar}, {Nakamura}, {Narayan}, {Narayanan}, {Natarajan}, {Neri}, {Ni}, {Noutsos}, {Okino}, {Olivares}, {Ortiz-Le{\'o}n}, {Oyama}, {{\"O}zel}, {Palumbo}, {Patel}, {Pen}, {Pesce}, {Pi{\'e}tu}, {Plambeck}, {PopStefanija}, {Porth}, {Prather}, {Preciado-L{\'o}pez}, {Psaltis}, {Pu}, {Ramakrishnan}, {Rao}, {Rawlings}, {Raymond}, {Rezzolla}, {Ripperda}, {Roelofs}, {Rogers}, {Ros}, {Rose}, {Roshanineshat}, {Rottmann}, {Roy}, {Ruszczyk}, {Ryan}, {Rygl}, {S{\'a}nchez}, {S{\'a}nchez-Arguelles}, {Sasada}, {Savolainen}, {Schloerb}, {Schuster}, {Shao}, {Shen}, {Small}, {Sohn}, {SooHoo}, {Tazaki}, {Tiede}, {Tilanus}, {Titus}, {Toma}, {Torne},
  {Trent}, {Trippe}, {Tsuda}, {van Bemmel}, {van Langevelde}, {van Rossum}, {Wagner}, {Wardle}, {Weintroub}, {Wex}, {Wharton}, {Wielgus}, {Wong}, {Wu}, {Young}, \& {Young}}]{eht2019}
{Event Horizon Telescope Collaboration}, {Akiyama}, K., {Alberdi}, A., {et~al.} 2019, \bibinfo{title}{{First M87 Event Horizon Telescope Results. I. The Shadow of the Supermassive Black Hole},} \apjl, 875, L1, \dodoi{10.3847/2041-8213/ab0ec7}

\bibitem[{ {Event Horizon Telescope Collaboration} {et~al.}(2022){Event Horizon Telescope Collaboration}, {Akiyama}, {Alberdi}, {Alef}, {Algaba}, {Anantua}, {Asada}, {Azulay}, {Bach}, {Baczko}, {Ball}, {Balokovi{\'c}}, {Barrett}, {Baub{\"o}ck}, {Benson}, {Bintley}, {Blackburn}, {Blundell}, {Bouman}, {Bower}, {Boyce}, {Bremer}, {Brinkerink}, {Brissenden}, {Britzen}, {Broderick}, {Broguiere}, {Bronzwaer}, {Bustamante}, {Byun}, {Carlstrom}, {Ceccobello}, {Chael}, {Chan}, {Chatterjee}, {Chatterjee}, {Chen}, {Chen}, {Cheng}, {Cho}, {Christian}, {Conroy}, {Conway}, {Cordes}, {Crawford}, {Crew}, {Cruz-Osorio}, {Cui}, {Davelaar}, {De Laurentis}, {Deane}, {Dempsey}, {Desvignes}, {Dexter}, {Dhruv}, {Doeleman}, {Dougal}, {Dzib}, {Eatough}, {Emami}, {Falcke}, {Farah}, {Fish}, {Fomalont}, {Ford}, {Fraga-Encinas}, {Freeman}, {Friberg}, {Fromm}, {Fuentes}, {Galison}, {Gammie}, {Garc{\'\i}a}, {Gentaz}, {Georgiev}, {Goddi}, {Gold}, {G{\'o}mez-Ruiz}, {G{\'o}mez}, {Gu}, {Gurwell}, {Hada}, {Haggard}, {Haworth}, {Hecht},
  {Hesper}, {Heumann}, {Ho}, {Ho}, {Honma}, {Huang}, {Huang}, {Hughes}, {Ikeda}, {Impellizzeri}, {Inoue}, {Issaoun}, {James}, {Jannuzi}, {Janssen}, {Jeter}, {Jiang}, {Jim{\'e}nez-Rosales}, {Johnson}, {Jorstad}, {Joshi}, {Jung}, {Karami}, {Karuppusamy}, {Kawashima}, {Keating}, {Kettenis}, {Kim}, {Kim}, {Kim}, {Kim}, {Kino}, {Koay}, {Kocherlakota}, {Kofuji}, {Koch}, {Koyama}, {Kramer}, {Kramer}, {Krichbaum}, {Kuo}, {La Bella}, {Lauer}, {Lee}, {Lee}, {Leung}, {Levis}, {Li}, {Lico}, {Lindahl}, {Lindqvist}, {Lisakov}, {Liu}, {Liu}, {Liuzzo}, {Lo}, {Lobanov}, {Loinard}, {Lonsdale}, {Lu}, {Mao}, {Marchili}, {Markoff}, {Marrone}, {Marscher}, {Mart{\'\i}-Vidal}, {Matsushita}, {Matthews}, {Medeiros}, {Menten}, {Michalik}, {Mizuno}, {Mizuno}, {Moran}, {Moriyama}, {Moscibrodzka}, {M{\"u}ller}, {Mus}, {Musoke}, {Myserlis}, {Nadolski}, {Nagai}, {Nagar}, {Nakamura}, {Narayan}, {Narayanan}, {Natarajan}, {Nathanail}, {Fuentes}, {Neilsen}, {Neri}, {Ni}, {Noutsos}, {Nowak}, {Oh}, {Okino}, {Olivares}, {Ortiz-Le{\'o}n}, {Oyama},
  {{\"O}zel}, {Palumbo}, {Paraschos}, {Park}, {Parsons}, {Patel}, {Pen}, {Pesce}, {Pi{\'e}tu}, {Plambeck}, {PopStefanija}, {Porth}, {P{\"o}tzl}, {Prather}, {Preciado-L{\'o}pez}, \& {Psaltis}}]{eht2022}
{Event Horizon Telescope Collaboration}, {Akiyama}, K., {Alberdi}, A., {et~al.} 2022, \bibinfo{title}{{First Sagittarius A* Event Horizon Telescope Results. I. The Shadow of the Supermassive Black Hole in the Center of the Milky Way},} \apjl, 930, L12, \dodoi{10.3847/2041-8213/ac6674}

\bibitem[{G. {Fonseca Alvarez} {et~al.}(2020){Fonseca Alvarez}, {Trump}, {Homayouni}, {Grier}, {Shen}, {Horne}, {Li}, {Brandt}, {Ho}, {Peterson}, \& {Schneider}}]{alvarez2020}
{Fonseca Alvarez}, G., {Trump}, J.~R., {Homayouni}, Y., {et~al.} 2020, \bibinfo{title}{{The Sloan Digital Sky Survey Reverberation Mapping Project: The H{\ensuremath{\beta}} Radius-Luminosity Relation},} \apj, 899, 73, \dodoi{10.3847/1538-4357/aba001}

\bibitem[{D. {Foreman-Mackey} {et~al.}(2013){Foreman-Mackey}, {Hogg}, {Lang}, \& {Goodman}}]{emcee}
{Foreman-Mackey}, D., {Hogg}, D.~W., {Lang}, D., \& {Goodman}, J. 2013, \bibinfo{title}{{emcee: The MCMC Hammer},} \pasp, 125, 306, \dodoi{10.1086/670067}

\bibitem[{G. {Ghisellini} {et~al.}(2014){Ghisellini}, {Tavecchio}, {Maraschi}, {Celotti}, \& {Sbarrato}}]{ghisellini2014}
{Ghisellini}, G., {Tavecchio}, F., {Maraschi}, L., {Celotti}, A., \& {Sbarrato}, T. 2014, \bibinfo{title}{{The power of relativistic jets is larger than the luminosity of their accretion disks},} \nat, 515, 376, \dodoi{10.1038/nature13856}

\bibitem[{ {Gravity Collaboration} {et~al.}(2018){Gravity Collaboration}, {Sturm}, {Dexter}, {Pfuhl}, {Stock}, {Davies}, {Lutz}, {Cl{\'e}net}, {Eckart}, {Eisenhauer}, {Genzel}, {Gratadour}, {H{\"o}nig}, {Kishimoto}, {Lacour}, {Millour}, {Netzer}, {Perrin}, {Peterson}, {Petrucci}, {Rouan}, {Waisberg}, {Woillez}, {Amorim}, {Brandner}, {F{\"o}rster Schreiber}, {Garcia}, {Gillessen}, {Ott}, {Paumard}, {Perraut}, {Scheithauer}, {Straubmeier}, {Tacconi}, \& {Widmann}}]{gravity2018}
{Gravity Collaboration}, {Sturm}, E., {Dexter}, J., {et~al.} 2018, \bibinfo{title}{{Spatially resolved rotation of the broad-line region of a quasar at sub-parsec scale},} \nat, 563, 657, \dodoi{10.1038/s41586-018-0731-9}

\bibitem[{P.~J. {Green} {et~al.}(2022){Green}, {Pulgarin-Duque}, {Anderson}, {MacLeod}, {Eracleous}, {Ruan}, {Runnoe}, {Graham}, {Roulston}, {Schneider}, {Ahlf}, {Bizyaev}, {Brownstein}, {del Casal}, {Dodd}, {Hoover}, {Matt}, {Merloni}, {Pan}, {Ramirez}, {Ridder}, \& {Moseley}}]{green2022}
{Green}, P.~J., {Pulgarin-Duque}, L., {Anderson}, S.~F., {et~al.} 2022, \bibinfo{title}{{The Time Domain Spectroscopic Survey: Changing-look Quasar Candidates from Multi-epoch Spectroscopy in SDSS-IV},} \apj, 933, 180, \dodoi{10.3847/1538-4357/ac743f}

\bibitem[{C.~J. {Grier} {et~al.}(2017){Grier}, {Trump}, {Shen}, {Horne}, {Kinemuchi}, {McGreer}, {Starkey}, {Brandt}, {Hall}, {Kochanek}, {Chen}, {Denney}, {Greene}, {Ho}, {Homayouni}, {I-Hsiu Li}, {Pei}, {Peterson}, {Petitjean}, {Schneider}, {Sun}, {AlSayyad}, {Bizyaev}, {Brinkmann}, {Brownstein}, {Bundy}, {Dawson}, {Eftekharzadeh}, {Fernandez-Trincado}, {Gao}, {Hutchinson}, {Jia}, {Jiang}, {Oravetz}, {Pan}, {Paris}, {Ponder}, {Peters}, {Rogerson}, {Simmons}, {Smith}, \& {Wang}}]{grier2017}
{Grier}, C.~J., {Trump}, J.~R., {Shen}, Y., {et~al.} 2017, \bibinfo{title}{{The Sloan Digital Sky Survey Reverberation Mapping Project: H{\ensuremath{\alpha}} and H{\ensuremath{\beta}} Reverberation Measurements from First-year Spectroscopy and Photometry},} \apj, 851, 21, \dodoi{10.3847/1538-4357/aa98dc}

\bibitem[{C.~J. {Grier} {et~al.}(2019){Grier}, {Shen}, {Horne}, {Brandt}, {Trump}, {Hall}, {Kinemuchi}, {Starkey}, {Schneider}, {Ho}, {Homayouni}, {I-Hsiu Li}, {McGreer}, {Peterson}, {Bizyaev}, {Chen}, {Dawson}, {Eftekharzadeh}, {Guo}, {Jia}, {Jiang}, {Kneib}, {Li}, {Li}, {Nie}, {Oravetz}, {Oravetz}, {Pan}, {Petitjean}, {Ponder}, {Rogerson}, {Vivek}, {Zhang}, \& {Zou}}]{grier2019}
{Grier}, C.~J., {Shen}, Y., {Horne}, K., {et~al.} 2019, \bibinfo{title}{{The Sloan Digital Sky Survey Reverberation Mapping Project: Initial C IV Lag Results from Four Years of Data},} \apj, 887, 38, \dodoi{10.3847/1538-4357/ab4ea5}

\bibitem[{H. {Guo} {et~al.}(2019){Guo}, {Liu}, {Shen}, {Loeb}, {Monroe}, \& {Prochaska}}]{guo2019}
{Guo}, H., {Liu}, X., {Shen}, Y., {et~al.} 2019, \bibinfo{title}{{Constraining sub-parsec binary supermassive black holes in quasars with multi-epoch spectroscopy - III. Candidates from continued radial velocity tests},} \mnras, 482, 3288, \dodoi{10.1093/mnras/sty2920}

\bibitem[{H. {Guo} {et~al.}(2020){Guo}, {Peng}, {Zhang}, {Burke}, {Liu}, {Sun}, {Wang}, {Kong}, {Sheng}, {Wang}, {He}, \& {Gu}}]{guo2020}
{Guo}, H., {Peng}, J., {Zhang}, K., {et~al.} 2020, \bibinfo{title}{{High-redshift Extreme Variability Quasars from Sloan Digital Sky Survey Multiepoch Spectroscopy},} \apj, 905, 52, \dodoi{10.3847/1538-4357/abc2ce}

\bibitem[{P.~C. {Hewett} {et~al.}(1995){Hewett}, {Foltz}, \& {Chaffee}}]{hewett1995}
{Hewett}, P.~C., {Foltz}, C.~B., \& {Chaffee}, F.~H. 1995, \bibinfo{title}{{The Large Bright Quasar Survey.VI.Quasar Catalog and Survey Parameters},} \aj, 109, 1498, \dodoi{10.1086/117380}

\bibitem[{Y. {Homayouni} {et~al.}(2019){Homayouni}, {Trump}, {Grier}, {Shen}, {Starkey}, {Brandt}, {Fonseca Alvarez}, {Hall}, {Horne}, {Kinemuchi}, {I-Hsiu Li}, {McGreer}, {Sun}, {Ho}, \& {Schneider}}]{homayouni2019}
{Homayouni}, Y., {Trump}, J.~R., {Grier}, C.~J., {et~al.} 2019, \bibinfo{title}{{The Sloan Digital Sky Survey Reverberation Mapping Project: Accretion Disk Sizes from Continuum Lags},} \apj, 880, 126, \dodoi{10.3847/1538-4357/ab2638}

\bibitem[{Y. {Homayouni} {et~al.}(2020){Homayouni}, {Trump}, {Grier}, {Horne}, {Shen}, {Brandt}, {Dawson}, {Alvarez}, {Green}, {Hall}, {Hern{\'a}ndez Santisteban}, {Ho}, {Kinemuchi}, {Kochanek}, {Li}, {Peterson}, {Schneider}, {Starkey}, {Bizyaev}, {Pan}, {Oravetz}, \& {Simmons}}]{homayouni2020}
{Homayouni}, Y., {Trump}, J.~R., {Grier}, C.~J., {et~al.} 2020, \bibinfo{title}{{The Sloan Digital Sky Survey Reverberation Mapping Project: Mg II Lag Results from Four Years of Monitoring},} \apj, 901, 55, \dodoi{10.3847/1538-4357/ababa9}

\bibitem[{B.~C. {Kelly} \& Y. {Shen}(2013){Kelly} \& {Shen}}]{kellyshen2013}
{Kelly}, B.~C., \& {Shen}, Y. 2013, \bibinfo{title}{{The Demographics of Broad-line Quasars in the Mass-Luminosity Plane. II. Black Hole Mass and Eddington Ratio Functions},} \apj, 764, 45, \dodoi{10.1088/0004-637X/764/1/45}

\bibitem[{J. {Kova{\v{c}}evi{\'c}} {et~al.}(2014){Kova{\v{c}}evi{\'c}}, {Popovi{\'c}}, \& {Kollatschny}}]{kovacevic2014}
{Kova{\v{c}}evi{\'c}}, J., {Popovi{\'c}}, L.~{\v{C}}., \& {Kollatschny}, W. 2014, \bibinfo{title}{{A model for the Balmer pseudocontinuum in spectra of type 1 AGNs},} Advances in Space Research, 54, 1347, \dodoi{10.1016/j.asr.2013.11.035}

\bibitem[{S. {Koz{\l}owski}(2016){Koz{\l}owski}}]{kozlowski2017}
{Koz{\l}owski}, S. 2016, \bibinfo{title}{{Revisiting Stochastic Variability of AGNs with Structure Functions},} \apj, 826, 118, \dodoi{10.3847/0004-637X/826/2/118}

\bibitem[{S.~M. {LaMassa} {et~al.}(2015){LaMassa}, {Cales}, {Moran}, {Myers}, {Richards}, {Eracleous}, {Heckman}, {Gallo}, \& {Urry}}]{lamassa2015}
{LaMassa}, S.~M., {Cales}, S., {Moran}, E.~C., {et~al.} 2015, \bibinfo{title}{{The Discovery of the First {\textquotedblleft}Changing Look{\textquotedblright} Quasar: New Insights Into the Physics and Phenomenology of Active Galactic Nucleus},} \apj, 800, 144, \dodoi{10.1088/0004-637X/800/2/144}

\bibitem[{M. {Lampton} {et~al.}(1976){Lampton}, {Margon}, \& {Bowyer}}]{lampton1976}
{Lampton}, M., {Margon}, B., \& {Bowyer}, S. 1976, \bibinfo{title}{{Parameter estimation in X-ray astronomy.},} \apj, 208, 177, \dodoi{10.1086/154592}

\bibitem[{A. {Laor} {et~al.}(2006){Laor}, {Barth}, {Ho}, \& {Filippenko}}]{laor2006}
{Laor}, A., {Barth}, A.~J., {Ho}, L.~C., \& {Filippenko}, A.~V. 2006, \bibinfo{title}{{Is the Broad-Line Region Clumped or Smooth? Constraints from the H{\ensuremath{\alpha}} Profile in NGC 4395, the Least Luminous Seyfert 1 Galaxy},} \apj, 636, 83, \dodoi{10.1086/497908}

\bibitem[{J.~I.~H. {Li} {et~al.}(2023){Li}, {Shen}, {Ho}, {Brandt}, {Grier}, {Hall}, {Homayouni}, {Koekemoer}, {Schneider}, \& {Trump}}]{li2023}
{Li}, J. I.~H., {Shen}, Y., {Ho}, L.~C., {et~al.} 2023, \bibinfo{title}{{The Sloan Digital Sky Survey Reverberation Mapping Project: The Black Hole Mass-Stellar Mass Relations at 0.2 {\ensuremath{\lesssim}} z {\ensuremath{\lesssim}} 0.8},} \apj, 954, 173, \dodoi{10.3847/1538-4357/acddda}

\bibitem[{B.~W. {Lyke} {et~al.}(2020){Lyke}, {Higley}, {McLane}, {Schurhammer}, {Myers}, {Ross}, {Dawson}, {Chabanier}, {Martini}, {Busca}, {Mas des Bourboux}, {Salvato}, {Streblyanska}, {Zarrouk}, {Burtin}, {Anderson}, {Bautista}, {Bizyaev}, {Brandt}, {Brinkmann}, {Brownstein}, {Comparat}, {Green}, {de la Macorra}, {Mu{\~n}oz Guti{\'e}rrez}, {Hou}, {Newman}, {Palanque-Delabrouille}, {P{\^a}ris}, {Percival}, {Petitjean}, {Rich}, {Rossi}, {Schneider}, {Smith}, {Vivek}, \& {Weaver}}]{lyke2020}
{Lyke}, B.~W., {Higley}, A.~N., {McLane}, J.~N., {et~al.} 2020, \bibinfo{title}{{The Sloan Digital Sky Survey Quasar Catalog: Sixteenth Data Release},} \apjs, 250, 8, \dodoi{10.3847/1538-4365/aba623}

\bibitem[{C.~L. {MacLeod} {et~al.}(2012){MacLeod}, {Ivezi{\'c}}, {Sesar}, {de Vries}, {Kochanek}, {Kelly}, {Becker}, {Lupton}, {Hall}, {Richards}, {Anderson}, \& {Schneider}}]{macleod2012}
{MacLeod}, C.~L., {Ivezi{\'c}}, {\v{Z}}., {Sesar}, B., {et~al.} 2012, \bibinfo{title}{{A Description of Quasar Variability Measured Using Repeated SDSS and POSS Imaging},} \apj, 753, 106, \dodoi{10.1088/0004-637X/753/2/106}

\bibitem[{C.~L. {MacLeod} {et~al.}(2016){MacLeod}, {Ross}, {Lawrence}, {Goad}, {Horne}, {Burgett}, {Chambers}, {Flewelling}, {Hodapp}, {Kaiser}, {Magnier}, {Wainscoat}, \& {Waters}}]{macleod2016}
{MacLeod}, C.~L., {Ross}, N.~P., {Lawrence}, A., {et~al.} 2016, \bibinfo{title}{{A systematic search for changing-look quasars in SDSS},} \mnras, 457, 389, \dodoi{10.1093/mnras/stv2997}

\bibitem[{C.~L. {MacLeod} {et~al.}(2018){MacLeod}, {Green}, {Anderson}, {Eracleous}, {Ruan}, {Runnoe}, {Brandt}, {Badenes}, {Greene}, {Morganson}, {Schmidt}, {Schwope}, {Shen}, {Amaro}, {Lebleu}, {Filiz Ak}, {Grier}, {Hoover}, {McGraw}, {Dawson}, {Hall}, {Hawley}, {Mariappan}, {Myers}, {P{\^a}ris}, {Schneider}, {Stassun}, {Bershady}, {Blanton}, {Seo}, {Tinker}, {Fern{\'a}ndez-Trincado}, {Chambers}, {Kaiser}, {Kudritzki}, {Magnier}, {Metcalfe}, \& {Waters}}]{macleod2018}
{MacLeod}, C.~L., {Green}, P.~J., {Anderson}, S.~F., {et~al.} 2018, \bibinfo{title}{{The Time-domain Spectroscopic Survey: Target Selection for Repeat Spectroscopy},} \aj, 155, 6, \dodoi{10.3847/1538-3881/aa99da}

\bibitem[{C.~L. {MacLeod} {et~al.}(2019){MacLeod}, {Green}, {Anderson}, {Bruce}, {Eracleous}, {Graham}, {Homan}, {Lawrence}, {LeBleu}, {Ross}, {Ruan}, {Runnoe}, {Stern}, {Burgett}, {Chambers}, {Kaiser}, {Magnier}, \& {Metcalfe}}]{macleod2019}
{MacLeod}, C.~L., {Green}, P.~J., {Anderson}, S.~F., {et~al.} 2019, \bibinfo{title}{{Changing-look Quasar Candidates: First Results from Follow-up Spectroscopy of Highly Optically Variable Quasars},} \apj, 874, 8, \dodoi{10.3847/1538-4357/ab05e2}

\bibitem[{E. {Morganson} {et~al.}(2015){Morganson}, {Green}, {Anderson}, {Ruan}, {Myers}, {Eracleous}, {Kelly}, {Badenes}, {Ba{\~n}ados}, {Blanton}, {Bershady}, {Borissova}, {Brandt}, {Burgett}, {Chambers}, {Draper}, {Davenport}, {Flewelling}, {Garnavich}, {Hawley}, {Hodapp}, {Isler}, {Kaiser}, {Kinemuchi}, {Kudritzki}, {Metcalfe}, {Morgan}, {P{\^a}ris}, {Parvizi}, {Poleski}, {Price}, {Salvato}, {Shanks}, {Schlafly}, {Schneider}, {Shen}, {Stassun}, {Tonry}, {Walter}, \& {Waters}}]{morganson2015}
{Morganson}, E., {Green}, P.~J., {Anderson}, S.~F., {et~al.} 2015, \bibinfo{title}{{The Time Domain Spectroscopic Survey: Variable Selection and Anticipated Results},} \apj, 806, 244, \dodoi{10.1088/0004-637X/806/2/244}

\bibitem[{I. {P{\^a}ris} {et~al.}(2012){P{\^a}ris}, {Petitjean}, {Aubourg}, {Bailey}, {Ross}, {Myers}, {Strauss}, {Anderson}, {Arnau}, {Bautista}, {Bizyaev}, {Bolton}, {Bovy}, {Brandt}, {Brewington}, {Browstein}, {Busca}, {Capellupo}, {Carithers}, {Croft}, {Dawson}, {Delubac}, {Ebelke}, {Eisenstein}, {Engelke}, {Fan}, {Filiz Ak}, {Finley}, {Font-Ribera}, {Ge}, {Gibson}, {Hall}, {Hamann}, {Hennawi}, {Ho}, {Hogg}, {Ivezi{\'c}}, {Jiang}, {Kimball}, {Kirkby}, {Kirkpatrick}, {Lee}, {Le Goff}, {Lundgren}, {MacLeod}, {Malanushenko}, {Malanushenko}, {Maraston}, {McGreer}, {McMahon}, {Miralda-Escud{\'e}}, {Muna}, {Noterdaeme}, {Oravetz}, {Palanque-Delabrouille}, {Pan}, {Perez-Fournon}, {Pieri}, {Richards}, {Rollinde}, {Sheldon}, {Schlegel}, {Schneider}, {Slosar}, {Shelden}, {Shen}, {Simmons}, {Snedden}, {Suzuki}, {Tinker}, {Viel}, {Weaver}, {Weinberg}, {White}, {Wood-Vasey}, \& {Y{\`e}che}}]{paris2012}
{P{\^a}ris}, I., {Petitjean}, P., {Aubourg}, {\'E}., {et~al.} 2012, \bibinfo{title}{{The Sloan Digital Sky Survey quasar catalog: ninth data release},} \aap, 548, A66, \dodoi{10.1051/0004-6361/201220142}

\bibitem[{I. {P{\^a}ris} {et~al.}(2014){P{\^a}ris}, {Petitjean}, {Aubourg}, {Ross}, {Myers}, {Streblyanska}, {Bailey}, {Hall}, {Strauss}, {Anderson}, {Bizyaev}, {Borde}, {Brinkmann}, {Bovy}, {Brandt}, {Brewington}, {Brownstein}, {Cook}, {Ebelke}, {Fan}, {Filiz Ak}, {Finley}, {Font-Ribera}, {Ge}, {Hamann}, {Ho}, {Jiang}, {Kinemuchi}, {Malanushenko}, {Malanushenko}, {Marchante}, {McGreer}, {McMahon}, {Miralda-Escud{\'e}}, {Muna}, {Noterdaeme}, {Oravetz}, {Palanque-Delabrouille}, {Pan}, {Perez-Fournon}, {Pieri}, {Riffel}, {Schlegel}, {Schneider}, {Simmons}, {Viel}, {Weaver}, {Wood-Vasey}, {Y{\`e}che}, \& {York}}]{paris2014}
{P{\^a}ris}, I., {Petitjean}, P., {Aubourg}, {\'E}., {et~al.} 2014, \bibinfo{title}{{The Sloan Digital Sky Survey quasar catalog: tenth data release},} \aap, 563, A54, \dodoi{10.1051/0004-6361/201322691}

\bibitem[{I. {P{\^a}ris} {et~al.}(2017){P{\^a}ris}, {Petitjean}, {Ross}, {Myers}, {Aubourg}, {Streblyanska}, {Bailey}, {Armengaud}, {Palanque-Delabrouille}, {Y{\`e}che}, {Hamann}, {Strauss}, {Albareti}, {Bovy}, {Bizyaev}, {Niel Brandt}, {Brusa}, {Buchner}, {Comparat}, {Croft}, {Dwelly}, {Fan}, {Font-Ribera}, {Ge}, {Georgakakis}, {Hall}, {Jiang}, {Kinemuchi}, {Malanushenko}, {Malanushenko}, {McMahon}, {Menzel}, {Merloni}, {Nandra}, {Noterdaeme}, {Oravetz}, {Pan}, {Pieri}, {Prada}, {Salvato}, {Schlegel}, {Schneider}, {Simmons}, {Viel}, {Weinberg}, \& {Zhu}}]{paris2017}
{P{\^a}ris}, I., {Petitjean}, P., {Ross}, N.~P., {et~al.} 2017, \bibinfo{title}{{The Sloan Digital Sky Survey Quasar Catalog: Twelfth data release},} \aap, 597, A79, \dodoi{10.1051/0004-6361/201527999}

\bibitem[{I. {P{\^a}ris} {et~al.}(2018){P{\^a}ris}, {Petitjean}, {Aubourg}, {Myers}, {Streblyanska}, {Lyke}, {Anderson}, {Armengaud}, {Bautista}, {Blanton}, {Blomqvist}, {Brinkmann}, {Brownstein}, {Brandt}, {Burtin}, {Dawson}, {de la Torre}, {Georgakakis}, {Gil-Mar{\'\i}n}, {Green}, {Hall}, {Kneib}, {LaMassa}, {Le Goff}, {MacLeod}, {Mariappan}, {McGreer}, {Merloni}, {Noterdaeme}, {Palanque-Delabrouille}, {Percival}, {Ross}, {Rossi}, {Schneider}, {Seo}, {Tojeiro}, {Weaver}, {Weijmans}, {Y{\`e}che}, {Zarrouk}, \& {Zhao}}]{paris2018}
{P{\^a}ris}, I., {Petitjean}, P., {Aubourg}, {\'E}., {et~al.} 2018, \bibinfo{title}{{The Sloan Digital Sky Survey Quasar Catalog: Fourteenth data release},} \aap, 613, A51, \dodoi{10.1051/0004-6361/201732445}

\bibitem[{B.~M. {Peterson} \& A. {Wandel}(2000){Peterson} \& {Wandel}}]{peterson2000}
{Peterson}, B.~M., \& {Wandel}, A. 2000, \bibinfo{title}{{Evidence for Supermassive Black Holes in Active Galactic Nuclei from Emission-Line Reverberation},} \apjl, 540, L13, \dodoi{10.1086/312862}

\bibitem[{B.~M. {Peterson} {et~al.}(1992){Peterson}, {Alloin}, {Axon}, {Balonek}, {Bertram}, {Boroson}, {Christensen}, {Clements}, {Dietrich}, {Elvis}, {Filippenko}, {Gaskell}, {Haswell}, {Huchra}, {Jackson}, {Kollatschny}, {Korista}, {Lame}, {Leacock}, {Lin}, {Malkan}, {Monk}, {Penston}, {Pogge}, {Robinson}, {Rosenblatt}, {Shields}, {Smith}, {Stirpe}, {Sun}, {Turner}, {Wagner}, {Wilkes}, \& {Wills}}]{peterson1992}
{Peterson}, B.~M., {Alloin}, D., {Axon}, D., {et~al.} 1992, \bibinfo{title}{{Steps toward Determination of the Size and Structure of the Broad-Line Region in Active Galactic Nuclei. III. Further Observations of NGC 5548 at Optical Wavelengths},} \apj, 392, 470, \dodoi{10.1086/171447}

\bibitem[{B. {Potts} \& C. {Villforth}(2021){Potts} \& {Villforth}}]{pottsvillforth2021}
{Potts}, B., \& {Villforth}, C. 2021, \bibinfo{title}{{A systematic search for changing-look quasars in SDSS-II using difference spectra},} \aap, 650, A33, \dodoi{10.1051/0004-6361/202140597}

\bibitem[{J.~C. {Runnoe} {et~al.}(2012){Runnoe}, {Brotherton}, \& {Shang}}]{runnoe2012}
{Runnoe}, J.~C., {Brotherton}, M.~S., \& {Shang}, Z. 2012, \bibinfo{title}{{Updating quasar bolometric luminosity corrections - II. Infrared bolometric corrections},} \mnras, 426, 2677, \dodoi{10.1111/j.1365-2966.2012.21644.x}

\bibitem[{J.~C. {Runnoe} {et~al.}(2015){Runnoe}, {Eracleous}, {Mathes}, {Pennell}, {Boroson}, {Sigur{\dh}sson}, {Bogdanovi{\'c}}, {Halpern}, \& {Liu}}]{runnoe2015}
{Runnoe}, J.~C., {Eracleous}, M., {Mathes}, G., {et~al.} 2015, \bibinfo{title}{{A Large Systematic Search for Close Supermassive Binary and Rapidly Recoiling Black Holes. II. Continued Spectroscopic Monitoring and Optical Flux Variability},} \apjs, 221, 7, \dodoi{10.1088/0067-0049/221/1/7}

\bibitem[{J.~C. {Runnoe} {et~al.}(2016){Runnoe}, {Cales}, {Ruan}, {Eracleous}, {Anderson}, {Shen}, {Green}, {Morganson}, {LaMassa}, {Greene}, {Dwelly}, {Schneider}, {Merloni}, {Georgakakis}, \& {Roman-Lopes}}]{runnoe2016}
{Runnoe}, J.~C., {Cales}, S., {Ruan}, J.~J., {et~al.} 2016, \bibinfo{title}{{Now you see it, now you don't: the disappearing central engine of the quasar J1011+5442},} \mnras, 455, 1691, \dodoi{10.1093/mnras/stv2385}

\bibitem[{J.~C. {Runnoe} {et~al.}(2017){Runnoe}, {Eracleous}, {Pennell}, {Mathes}, {Boroson}, {Sigur{\dh}sson}, {Bogdanovi{\'c}}, {Halpern}, {Liu}, \& {Brown}}]{runnoe2017}
{Runnoe}, J.~C., {Eracleous}, M., {Pennell}, A., {et~al.} 2017, \bibinfo{title}{{A large systematic search for close supermassive binary and rapidly recoiling black holes - III. Radial velocity variations},} \mnras, 468, 1683, \dodoi{10.1093/mnras/stx452}

\bibitem[{P. {S{\'a}nchez-Bl{\'a}zquez} {et~al.}(2006){S{\'a}nchez-Bl{\'a}zquez}, {Peletier}, {Jim{\'e}nez-Vicente}, {Cardiel}, {Cenarro}, {Falc{\'o}n-Barroso}, {Gorgas}, {Selam}, \& {Vazdekis}}]{miles}
{S{\'a}nchez-Bl{\'a}zquez}, P., {Peletier}, R.~F., {Jim{\'e}nez-Vicente}, J., {et~al.} 2006, \bibinfo{title}{{Medium-resolution Isaac Newton Telescope library of empirical spectra},} \mnras, 371, 703, \dodoi{10.1111/j.1365-2966.2006.10699.x}

\bibitem[{M. {Schmidt} \& R.~F. {Green}(1983){Schmidt} \& {Green}}]{schmidtgreen1983}
{Schmidt}, M., \& {Green}, R.~F. 1983, \bibinfo{title}{{Quasar evolution derived from the Palomar bright quasar survey and other complete quasar surveys.},} \apj, 269, 352, \dodoi{10.1086/161048}

\bibitem[{D.~P. {Schneider} {et~al.}(2003){Schneider}, {Fan}, {Hall}, {Jester}, {Richards}, {Stoughton}, {Strauss}, {SubbaRao}, {Vanden Berk}, {Anderson}, {Brandt}, {Gunn}, {Gray}, {Trump}, {Voges}, {Yanny}, {Bahcall}, {Blanton}, {Boroski}, {Brinkmann}, {Brunner}, {Burles}, {Castander}, {Doi}, {Eisenstein}, {Frieman}, {Fukugita}, {Heckman}, {Hennessy}, {Ivezi{\'c}}, {Kent}, {Knapp}, {Lamb}, {Lee}, {Loveday}, {Lupton}, {Margon}, {Meiksin}, {Munn}, {Newberg}, {Nichol}, {Niederste-Ostholt}, {Pier}, {Richmond}, {Rockosi}, {Saxe}, {Schlegel}, {Szalay}, {Thakar}, {Uomoto}, \& {York}}]{schneider2003}
{Schneider}, D.~P., {Fan}, X., {Hall}, P.~B., {et~al.} 2003, \bibinfo{title}{{The Sloan Digital Sky Survey Quasar Catalog. II. First Data Release},} \aj, 126, 2579, \dodoi{10.1086/379174}

\bibitem[{D.~P. {Schneider} {et~al.}(2005){Schneider}, {Hall}, {Richards}, {Vanden Berk}, {Anderson}, {Fan}, {Jester}, {Stoughton}, {Strauss}, {SubbaRao}, {Brandt}, {Gunn}, {Yanny}, {Bahcall}, {Barentine}, {Blanton}, {Boroski}, {Brewington}, {Brinkmann}, {Brunner}, {Csabai}, {Doi}, {Eisenstein}, {Frieman}, {Fukugita}, {Gray}, {Harvanek}, {Heckman}, {Ivezi{\'c}}, {Kent}, {Kleinman}, {Knapp}, {Kron}, {Krzesinski}, {Long}, {Loveday}, {Lupton}, {Margon}, {Munn}, {Neilsen}, {Newberg}, {Newman}, {Nichol}, {Nitta}, {Pier}, {Rockosi}, {Saxe}, {Schlegel}, {Snedden}, {Szalay}, {Thakar}, {Uomoto}, {Voges}, \& {York}}]{schneider2005}
{Schneider}, D.~P., {Hall}, P.~B., {Richards}, G.~T., {et~al.} 2005, \bibinfo{title}{{The Sloan Digital Sky Survey Quasar Catalog. III. Third Data Release},} \aj, 130, 367, \dodoi{10.1086/431156}

\bibitem[{D.~P. {Schneider} {et~al.}(2007){Schneider}, {Hall}, {Richards}, {Strauss}, {Vanden Berk}, {Anderson}, {Brandt}, {Fan}, {Jester}, {Gray}, {Gunn}, {SubbaRao}, {Thakar}, {Stoughton}, {Szalay}, {Yanny}, {York}, {Bahcall}, {Barentine}, {Blanton}, {Brewington}, {Brinkmann}, {Brunner}, {Castander}, {Csabai}, {Frieman}, {Fukugita}, {Harvanek}, {Hogg}, {Ivezi{\'c}}, {Kent}, {Kleinman}, {Knapp}, {Kron}, {Krzesi{\'n}ski}, {Long}, {Lupton}, {Nitta}, {Pier}, {Saxe}, {Shen}, {Snedden}, {Weinberg}, \& {Wu}}]{schneider2007}
{Schneider}, D.~P., {Hall}, P.~B., {Richards}, G.~T., {et~al.} 2007, \bibinfo{title}{{The Sloan Digital Sky Survey Quasar Catalog. IV. Fifth Data Release},} \aj, 134, 102, \dodoi{10.1086/518474}

\bibitem[{D.~P. {Schneider} {et~al.}(2010){Schneider}, {Richards}, {Hall}, {Strauss}, {Anderson}, {Boroson}, {Ross}, {Shen}, {Brandt}, {Fan}, {Inada}, {Jester}, {Knapp}, {Krawczyk}, {Thakar}, {Vanden Berk}, {Voges}, {Yanny}, {York}, {Bahcall}, {Bizyaev}, {Blanton}, {Brewington}, {Brinkmann}, {Eisenstein}, {Frieman}, {Fukugita}, {Gray}, {Gunn}, {Hibon}, {Ivezi{\'c}}, {Kent}, {Kron}, {Lee}, {Lupton}, {Malanushenko}, {Malanushenko}, {Oravetz}, {Pan}, {Pier}, {Price}, {Saxe}, {Schlegel}, {Simmons}, {Snedden}, {SubbaRao}, {Szalay}, \& {Weinberg}}]{schneider2010}
{Schneider}, D.~P., {Richards}, G.~T., {Hall}, P.~B., {et~al.} 2010, \bibinfo{title}{{The Sloan Digital Sky Survey Quasar Catalog. V. Seventh Data Release},} \aj, 139, 2360, \dodoi{10.1088/0004-6256/139/6/2360}

\bibitem[{S.~G. {Sergeev} {et~al.}(2007){Sergeev}, {Doroshenko}, {Dzyuba}, {Peterson}, {Pogge}, \& {Pronik}}]{sergeev2007}
{Sergeev}, S.~G., {Doroshenko}, V.~T., {Dzyuba}, S.~A., {et~al.} 2007, \bibinfo{title}{{Thirty Years of Continuum and Emission-Line Variability in NGC 5548},} \apj, 668, 708, \dodoi{10.1086/520697}

\bibitem[{R.~O. {Sexton} {et~al.}(2021){Sexton}, {Matzko}, {Darden}, {Canalizo}, \& {Gorjian}}]{badass}
{Sexton}, R.~O., {Matzko}, W., {Darden}, N., {Canalizo}, G., \& {Gorjian}, V. 2021, \bibinfo{title}{{Bayesian AGN Decomposition Analysis for SDSS spectra: a correlation analysis of [O III] {\ensuremath{\lambda}}5007 outflow kinematics with AGN and host galaxy properties},} \mnras, 500, 2871, \dodoi{10.1093/mnras/staa3278}

\bibitem[{Y. {Shen} \& L.~C. {Ho}(2014){Shen} \& {Ho}}]{shen2014nature}
{Shen}, Y., \& {Ho}, L.~C. 2014, \bibinfo{title}{{The diversity of quasars unified by accretion and orientation},} \nat, 513, 210, \dodoi{10.1038/nature13712}

\bibitem[{Y. {Shen} {et~al.}(2013){Shen}, {Liu}, {Loeb}, \& {Tremaine}}]{shen2013}
{Shen}, Y., {Liu}, X., {Loeb}, A., \& {Tremaine}, S. 2013, \bibinfo{title}{{Constraining Sub-parsec Binary Supermassive Black Holes in Quasars with Multi-epoch Spectroscopy. I. The General Quasar Population},} \apj, 775, 49, \dodoi{10.1088/0004-637X/775/1/49}

\bibitem[{Y. {Shen} \&  {SDSS-RM Collaboration}(2016){Shen} \& {SDSS-RM Collaboration}}]{shen2016}
{Shen}, Y., \& {SDSS-RM Collaboration}. 2016, \bibinfo{title}{{SDSS-RM: A Multi-Object AGN Reverberation Mapping Project},} in Astronomical Society of the Pacific Conference Series, Vol. 507, Multi-Object Spectroscopy in the Next Decade: Big Questions, Large Surveys, and Wide Fields, ed. I.~{Skillen}, M.~{Balcells}, \& S.~{Trager}, 367

\bibitem[{Y. {Shen} {et~al.}(2011){Shen}, {Richards}, {Strauss}, {Hall}, {Schneider}, {Snedden}, {Bizyaev}, {Brewington}, {Malanushenko}, {Malanushenko}, {Oravetz}, {Pan}, \& {Simmons}}]{shen2011}
{Shen}, Y., {Richards}, G.~T., {Strauss}, M.~A., {et~al.} 2011, \bibinfo{title}{{A Catalog of Quasar Properties from Sloan Digital Sky Survey Data Release 7},} \apjs, 194, 45, \dodoi{10.1088/0067-0049/194/2/45}

\bibitem[{Y. {Shen} {et~al.}(2015){Shen}, {Brandt}, {Dawson}, {Hall}, {McGreer}, {Anderson}, {Chen}, {Denney}, {Eftekharzadeh}, {Fan}, {Gao}, {Green}, {Greene}, {Ho}, {Horne}, {Jiang}, {Kelly}, {Kinemuchi}, {Kochanek}, {P{\^a}ris}, {Peters}, {Peterson}, {Petitjean}, {Ponder}, {Richards}, {Schneider}, {Seth}, {Smith}, {Strauss}, {Tao}, {Trump}, {Wood-Vasey}, {Zu}, {Eisenstein}, {Pan}, {Bizyaev}, {Malanushenko}, {Malanushenko}, \& {Oravetz}}]{rm}
{Shen}, Y., {Brandt}, W.~N., {Dawson}, K.~S., {et~al.} 2015, \bibinfo{title}{{The Sloan Digital Sky Survey Reverberation Mapping Project: Technical Overview},} \apjs, 216, 4, \dodoi{10.1088/0067-0049/216/1/4}

\bibitem[{Y. {Shen} {et~al.}(2019){Shen}, {Hall}, {Horne}, {Zhu}, {McGreer}, {Simm}, {Trump}, {Kinemuchi}, {Brandt}, {Green}, {Grier}, {Guo}, {Ho}, {Homayouni}, {Jiang}, {I-Hsiu Li}, {Morganson}, {Petitjean}, {Richards}, {Schneider}, {Starkey}, {Wang}, {Chambers}, {Kaiser}, {Kudritzki}, {Magnier}, \& {Waters}}]{shen2019}
{Shen}, Y., {Hall}, P.~B., {Horne}, K., {et~al.} 2019, \bibinfo{title}{{The Sloan Digital Sky Survey Reverberation Mapping Project: Sample Characterization},} \apjs, 241, 34, \dodoi{10.3847/1538-4365/ab074f}

\bibitem[{Y. {Shen} {et~al.}(2023){Shen}, {Grier}, {Horne}, {Stone}, {Li}, {Yang}, {Homayouni}, {Trump}, {Anderson}, {Brandt}, {Hall}, {Ho}, {Jiang}, {Petitjean}, {Schneider}, {Tao}, {Donnan}, {AlSayyad}, {Bershady}, {Blanton}, {Bizyaev}, {Bundy}, {Chen}, {Davis}, {Dawson}, {Fan}, {Greene}, {Groller}, {Guo}, {Ibarra-Medel}, {Jiang}, {Keenan}, {Kollmeier}, {Lejoly}, {Li}, {de la Macorra}, {Moe}, {Nie}, {Rossi}, {Smith}, {Tee}, {Weijmans}, {Xu}, {Yue}, {Zhou}, {Zhou}, \& {Zou}}]{shen2023}
{Shen}, Y., {Grier}, C.~J., {Horne}, K., {et~al.} 2023, \bibinfo{title}{{The Sloan Digital Sky Survey Reverberation Mapping Project: Key Results},} arXiv e-prints, arXiv:2305.01014, \dodoi{10.48550/arXiv.2305.01014}

\bibitem[{Y. {Shen} {et~al.}(2024){Shen}, {Grier}, {Horne}, {Stone}, {Li}, {Yang}, {Homayouni}, {Trump}, {Anderson}, {Brandt}, {Hall}, {Ho}, {Jiang}, {Petitjean}, {Schneider}, {Tao}, {Donnan}, {AlSayyad}, {Bershady}, {Blanton}, {Bizyaev}, {Bundy}, {Chen}, {Davis}, {Dawson}, {Fan}, {Greene}, {Gr{\"o}ller}, {Guo}, {Ibarra-Medel}, {Jiang}, {Keenan}, {Kollmeier}, {Lejoly}, {Li}, {de la Macorra}, {Moe}, {Nie}, {Rossi}, {Smith}, {Tee}, {Weijmans}, {Xu}, {Yue}, {Zhou}, {Zhou}, \& {Zou}}]{shen2024}
{Shen}, Y., {Grier}, C.~J., {Horne}, K., {et~al.} 2024, \bibinfo{title}{{The Sloan Digital Sky Survey Reverberation Mapping Project: Key Results},} \apjs, 272, 26, \dodoi{10.3847/1538-4365/ad3936}

\bibitem[{S.~A. {Smee} {et~al.}(2013){Smee}, {Gunn}, {Uomoto}, {Roe}, {Schlegel}, {Rockosi}, {Carr}, {Leger}, {Dawson}, {Olmstead}, {Brinkmann}, {Owen}, {Barkhouser}, {Honscheid}, {Harding}, {Long}, {Lupton}, {Loomis}, {Anderson}, {Annis}, {Bernardi}, {Bhardwaj}, {Bizyaev}, {Bolton}, {Brewington}, {Briggs}, {Burles}, {Burns}, {Castander}, {Connolly}, {Davenport}, {Ebelke}, {Epps}, {Feldman}, {Friedman}, {Frieman}, {Heckman}, {Hull}, {Knapp}, {Lawrence}, {Loveday}, {Mannery}, {Malanushenko}, {Malanushenko}, {Merrelli}, {Muna}, {Newman}, {Nichol}, {Oravetz}, {Pan}, {Pope}, {Ricketts}, {Shelden}, {Sandford}, {Siegmund}, {Simmons}, {Smith}, {Snedden}, {Schneider}, {SubbaRao}, {Tremonti}, {Waddell}, \& {York}}]{smee2013}
{Smee}, S.~A., {Gunn}, J.~E., {Uomoto}, A., {et~al.} 2013, \bibinfo{title}{{The Multi-object, Fiber-fed Spectrographs for the Sloan Digital Sky Survey and the Baryon Oscillation Spectroscopic Survey},} \aj, 146, 32, \dodoi{10.1088/0004-6256/146/2/32}

\bibitem[{P.~J. {Storey} \& D.~G. {Hummer}(1995){Storey} \& {Hummer}}]{storey1995}
{Storey}, P.~J., \& {Hummer}, D.~G. 1995, \bibinfo{title}{{Recombination line intensities for hydrogenic ions-IV. Total recombination coefficients and machine-readable tables for Z=1 to 8},} \mnras, 272, 41, \dodoi{10.1093/mnras/272.1.41}

\bibitem[{I. {Wanders} \& B.~M. {Peterson}(1996){Wanders} \& {Peterson}}]{wanders1996}
{Wanders}, I., \& {Peterson}, B.~M. 1996, \bibinfo{title}{{A Long-Term Study of Broad Emission Line Profile Variability in NGC 5548},} \apj, 466, 174, \dodoi{10.1086/177502}

\bibitem[{S. {Wang} {et~al.}(2020){Wang}, {Shen}, {Jiang}, {Grier}, {Horne}, {Homayouni}, {Peterson}, {Trump}, {Brandt}, {Hall}, {Ho}, {Li}, {Hernandez Santisteban}, {Kinemuchi}, {McGreer}, \& {Schneider}}]{wang2020}
{Wang}, S., {Shen}, Y., {Jiang}, L., {et~al.} 2020, \bibinfo{title}{{The Sloan Digital Sky Survey Reverberation Mapping Project: How Broad Emission Line Widths Change When Luminosity Changes},} \apj, 903, 51, \dodoi{10.3847/1538-4357/abb36d}

\bibitem[{B.~J. {Wills} \& I.~W.~A. {Browne}(1986){Wills} \& {Browne}}]{wills&browne1986}
{Wills}, B.~J., \& {Browne}, I.~W.~A. 1986, \bibinfo{title}{{Relativistic Beaming and Quasar Emission Lines},} \apj, 302, 56, \dodoi{10.1086/163973}

\bibitem[{Q. {Wu} \& Y. {Shen}(2022){Wu} \& {Shen}}]{wushen2022}
{Wu}, Q., \& {Shen}, Y. 2022, \bibinfo{title}{{A Catalog of Quasar Properties from Sloan Digital Sky Survey Data Release 16},} \apjs, 263, 42, \dodoi{10.3847/1538-4365/ac9ead}

\bibitem[{Q. {Yang} {et~al.}(2018){Yang}, {Wu}, {Fan}, {Jiang}, {McGreer}, {Shangguan}, {Yao}, {Wang}, {Joshi}, {Green}, {Wang}, {Feng}, {Fu}, {Yang}, \& {Liu}}]{yang2018}
{Yang}, Q., {Wu}, X.-B., {Fan}, X., {et~al.} 2018, \bibinfo{title}{{Discovery of 21 New Changing-look AGNs in the Northern Sky},} \apj, 862, 109, \dodoi{10.3847/1538-4357/aaca3a}

\bibitem[{D.~G. {York} {et~al.}(2000){York}, {Adelman}, {Anderson}, {Anderson}, {Annis}, {Bahcall}, {Bakken}, {Barkhouser}, {Bastian}, {Berman}, {Boroski}, {Bracker}, {Briegel}, {Briggs}, {Brinkmann}, {Brunner}, {Burles}, {Carey}, {Carr}, {Castander}, {Chen}, {Colestock}, {Connolly}, {Crocker}, {Csabai}, {Czarapata}, {Davis}, {Doi}, {Dombeck}, {Eisenstein}, {Ellman}, {Elms}, {Evans}, {Fan}, {Federwitz}, {Fiscelli}, {Friedman}, {Frieman}, {Fukugita}, {Gillespie}, {Gunn}, {Gurbani}, {de Haas}, {Haldeman}, {Harris}, {Hayes}, {Heckman}, {Hennessy}, {Hindsley}, {Holm}, {Holmgren}, {Huang}, {Hull}, {Husby}, {Ichikawa}, {Ichikawa}, {Ivezi{\'c}}, {Kent}, {Kim}, {Kinney}, {Klaene}, {Kleinman}, {Kleinman}, {Knapp}, {Korienek}, {Kron}, {Kunszt}, {Lamb}, {Lee}, {Leger}, {Limmongkol}, {Lindenmeyer}, {Long}, {Loomis}, {Loveday}, {Lucinio}, {Lupton}, {MacKinnon}, {Mannery}, {Mantsch}, {Margon}, {McGehee}, {McKay}, {Meiksin}, {Merelli}, {Monet}, {Munn}, {Narayanan}, {Nash}, {Neilsen}, {Neswold}, {Newberg}, {Nichol},
  {Nicinski}, {Nonino}, {Okada}, {Okamura}, {Ostriker}, {Owen}, {Pauls}, {Peoples}, {Peterson}, {Petravick}, {Pier}, {Pope}, {Pordes}, {Prosapio}, {Rechenmacher}, {Quinn}, {Richards}, {Richmond}, {Rivetta}, {Rockosi}, {Ruthmansdorfer}, {Sandford}, {Schlegel}, {Schneider}, {Sekiguchi}, {Sergey}, {Shimasaku}, {Siegmund}, {Smee}, {Smith}, {Snedden}, {Stone}, {Stoughton}, {Strauss}, {Stubbs}, {SubbaRao}, {Szalay}, {Szapudi}, {Szokoly}, {Thakar}, {Tremonti}, {Tucker}, {Uomoto}, {Vanden Berk}, {Vogeley}, {Waddell}, {Wang}, {Watanabe}, {Weinberg}, {Yanny}, {Yasuda}, \& {SDSS Collaboration}}]{york2000}
{York}, D.~G., {Adelman}, J., {Anderson}, John~E., J., {et~al.} 2000, \bibinfo{title}{{The Sloan Digital Sky Survey: Technical Summary},} \aj, 120, 1579, \dodoi{10.1086/301513}

\bibitem[{G. {Zeltyn} {et~al.}(2022){Zeltyn}, {Trakhtenbrot}, {Eracleous}, {Runnoe}, {Trump}, {Stern}, {Shen}, {Hern{\'a}ndez-Garc{\'\i}a}, {Bauer}, {Yang}, {Dwelly}, {Ricci}, {Green}, {Anderson}, {Assef}, {Guolo}, {MacLeod}, {Davis}, {Fries}, {Gezari}, {Grogin}, {Homan}, {Koekemoer}, {Krumpe}, {LaMassa}, {Liu}, {Merloni}, {Mart{\'\i}nez-Aldama}, {Schneider}, {Temple}, {Brownstein}, {Ibarra-Medel}, {Burke}, {Pellegrino}, \& {Kollmeier}}]{zeltyn2022}
{Zeltyn}, G., {Trakhtenbrot}, B., {Eracleous}, M., {et~al.} 2022, \bibinfo{title}{{A Transient ``Changing-look'' Active Galactic Nucleus Resolved on Month Timescales from First-year Sloan Digital Sky Survey V Data},} \apjl, 939, L16, \dodoi{10.3847/2041-8213/ac9a47}

\bibitem[{G. {Zeltyn} {et~al.}(2024){Zeltyn}, {Trakhtenbrot}, {Eracleous}, {Yang}, {Green}, {Anderson}, {LaMassa}, {Runnoe}, {Assef}, {Bauer}, {Brandt}, {Davis}, {Frederick}, {Fries}, {Graham}, {Grogin}, {Guolo}, {Hern{\'a}ndez-Garc{\'\i}a}, {Koekemoer}, {Krumpe}, {Liu}, {Mart{\'\i}nez-Aldama}, {Ricci}, {Schneider}, {Shen}, {{\'S}niegowska}, {Temple}, {Trump}, {Xue}, {Brownstein}, {Dwelly}, {Morrison}, {Bizyaev}, {Pan}, \& {Kollmeier}}]{zeltyn2024}
{Zeltyn}, G., {Trakhtenbrot}, B., {Eracleous}, M., {et~al.} 2024, \bibinfo{title}{{Exploring Changing-look Active Galactic Nuclei with the Sloan Digital Sky Survey V: First Year Results},} \apj, 966, 85, \dodoi{10.3847/1538-4357/ad2f30}

\end{thebibliography}
\bibliographystyle{aasjournalv7.bst}
\end{document}